\documentstyle[epsfig,12pt,aps]{revtex}
\begin{document}
\newcommand{\p}{\partial}
\newcommand{\D}{\Delta}
\newcommand{\ls}{\left(}
\newcommand{\rs}{\right)}
\newcommand{\beq}{\begin{equation}}
\newcommand{\eeq}{\end{equation}}
\newcommand{\beqa}{\begin{eqnarray}}
\newcommand{\eeqa}{\end{eqnarray}}
\newcommand{\bdm}{\begin{displaymath}}
\newcommand{\edm}{\end{displaymath}}
\newcommand{\fps}{f_{\pi}^2 }
\newcommand{\mks}{m_{{\mathrm K}}^2 }
\newcommand{\ms}{m_{{\mathrm K}}^{*} }
\newcommand{\mk}{m_{{\mathrm K}} }
\newcommand{\msq}{m_{{\mathrm K}}^{*2} }
\newcommand{\rhos}{\rho_{\mathrm s} }
\newcommand{\rhob}{\rho_{\mathrm B} }
\title{Dilepton production in heavy ion collisions at 
intermediate energies}
\author{K. Shekhter$^a$,  C. Fuchs$^a$, Amand Faessler$^a$, M. Krivoruchenko$^{a,b}$, 
B. Martemyanov$^{a,b}$}
\address{$^a$ Institut f\"ur Theoretische Physik der Universit\"at T\"ubingen, 
Auf der Morgenstelle 14, D-72076 T\"ubingen, Germany}
\address{$^b$ Institute for Theoretical and Experimental
Physics, B. Cheremushkinskaya 25, 117259 Moscow, Russia}
\maketitle  
\vspace{1cm}
\begin{abstract}
We present a unified description of the vector 
meson and dilepton production in elementary 
and in heavy ion reactions. The production of vector mesons ($\rho,\omega$) 
is described via the excitation of nuclear resonances ($R$). 
The theoretical framework is an extended vector meson dominance 
model (eVMD). The treatment of the resonance decays $R\longmapsto NV$ with 
arbitrary spin is covariant and kinematically complete. The eVMD 
includes thereby excited vector meson states  in the 
transition form factors. This ensures correct asymptotics and provides a 
unified description of photonic and mesonic decays. The resonance model 
is successfully applied to the $\omega$ production in $p+p$ reactions. 
The same model is applied to the 
dilepton production in elementary reactions ($p+p, p+d$). Corresponding 
data are well reproduced. However, when the model is applied to 
heavy ion reactions in the BEVALAC/SIS energy range  
the experimental dilepton spectra measured by the DLS Collaboration 
are significantly underestimated at small invariant masses. 
As a possible solution of this problem the destruction of 
quantum interference in a dense medium is discussed. A decoherent 
emission through vector mesons decays enhances the corresponding 
dilepton yield in heavy ion reactions. In the vicinity of the 
$\rho/\omega$-peak the reproduction of the data requires further 
a substantial collisional 
broadening of the $\rho$ and in particular of the $\omega$ meson.
\end{abstract}

\section{Introduction}
One of the important questions which theorists face at present 
is the dependence of hadron properties on medium effects. 
Medium effects manifest themselves in the modification of 
widths and masses of resonances produced in nuclear collisions.
The magnitude of such changes depends thereby on the density and 
the temperature of the medium. 
E.g., the proposed Brown-Rho scaling \cite{BR} is equivalent to a reduction 
of the vector meson masses in the nuclear medium. The same conclusion 
is obtained from QCD sum rules \cite{QCDSR}
and within effective hadronic models \cite{BGP}. The dispersion 
analysis of forward scattering amplitudes 
\cite{KKW,EIoffe,EBEK,KS} showed that vector meson mass shifts are in 
general small and positive, whereas at low momenta they can change 
the sign which is in qualitative agreement with the 
Brown-Rho scaling and the results from QCD sum rules. However, 
the question of in-medium masses must finally 
be settled experimentally. 

Dilepton spectra from heavy-ion collisions are considered 
as a suitable tool for this
purpose. The CERES \cite{ceres} and HELIOS \cite{HELIOS} 
Collaborations measured dilepton spectra at CERN and found
a significant enhancement of the low-energy dilepton yield below the 
$\rho $ and $\omega $ peaks \cite{ceres} in heavy reaction systems 
($Pb+Au$) compared to light systems ($S+W$) and 
proton induced reactions ($p+Be$). Theoretically, this enhancement 
can be explained within a hadronic picture by the assumption of 
a dropping $\rho $ mass \cite{drop} or by the inclusion 
of in-medium spectral functions for the vector mesons 
\cite{rapp,BCRW98}. In both cases the enhanced low energetic 
dilepton yield is not simply caused by a shift of the 
$\rho $ and $\omega $ peaks in the nuclear medium but it originates 
to most extent from an enhanced contribution of the $\pi^+ \pi^-$ annihilation 
channel which, assuming vector dominance, runs over an intermediate 
$\rho$ meson. An alternative scenario could be the formation 
of a quark-gluon plasma which leads to additional 
($pQCD$) contributions to the dilepton spectrum 
\cite{rapp,weise00}.

A similar situation occurs at a completely different energy scale,
 namely around 1 A.GeV incident energies where the low mass region
of dilepton spectra are underestimated by present transport
calculations compared to $pp$ and $pd$ reactions. The corresponding 
data were obtained by the DLS Collaboration at the BEVALAC \cite{DLS}. 
However, in contrast to ultra-relativistic reactions (SPS) 
the situation does not improve when full spectral functions
and/or a dropping mass of the vector mesons are taken
into account \cite{ernst,BK,BCRW98}. This fact is known as 
the DLS {\it puzzle}. The reason lies in the fact 
that both, possible $pQCD$ contributions as well as a sufficient amount of 
$\pi^+ \pi^-$ annihilation processes are absent at intermediate energies. 
Also a dropping $\eta$ mass 
can be excluded as a possible explanation of the DLS puzzle since it 
would contradict $m_T$ scaling \cite{BCRW98}. Furthermore, chiral 
perturbation theory predicts only very small modifications of the 
in-medium $\eta$ mass \cite{oset02}. 
Thus one has to search for other sources which could explain the low mass 
dilepton excess seen in heavy ion reactions. 
Dilepton spectra were also measured at KEK in $p + A$ reactions at a beam 
energy of 12 GeV \cite{KEK}. Also here an excess of 
dileptons compared to the known sources was 
observed below the $\rho$-meson peak and interpreted as a change of the 
 vector meson spectral functions. These data were recently analyzed in 
Ref. \cite{Elena},  again without success to explain the experimental 
spectrum within a  dropping mass scenario and/or by a 
significant collision broadening
of the vector mesons. Since the vector meson peaks are not 
resolved experimentally \cite{DLS}, the problem to extract in-medium 
masses directly from experimental data remains extremely difficult. 

For all these studies a precise and rather complete knowledge of the relative
weights for existing decay channels is indispensable in order to
draw reliable conclusions from dilepton spectra. 
In \cite{krivo00} a systematic study of meson decay channels
was performed, including channels which have been neglected so far, 
such as e.g. four-body decays $\rho^0\rightarrow\pi^0\pi^0e^+e^-$. 
However, as has been shown in \cite{resdec} in $pp$ reactions 
the contributions of these more exotic channels are not large enough 
to enhance the low mass dilepton yield at incident energies around 
1 AGeV. Here the low mass dilepton spectrum is dominated by the $\eta$ 
and the contributions from the decay of baryonic resonances 
\cite{ernst,resdec,BCEM}. 

The importance of the resonance contribution to the dilepton 
yield in elementary and heavy ion reactions has been stressed in several works 
\cite{resdec,koch96,pirner,post01,titov,zetenyi,BCM,resgiessen2,lutz03,mosel03,krivo01,krivo02}.  
In \cite{krivo02} we calculated in a fully relativistic 
treatment of the dilepton decays $R\rightarrow N\ e^+\ e^-$ of nucleon 
resonances with masses below 2 GeV. Kinematically complete 
phenomenological expressions for the dilepton decays of 
resonances with arbitrary spin and parity,
parameterized in terms of the magnetic, electric, and Coulomb transition form
factors and numerical estimates for the dilepton spectra and branching
ratios of the nucleon resonances were given. In \cite{resdec} this approach 
was applied to the dilepton production in $pp$ reactions at BEVALAC 
energies. In Sect. II. the theoretical framework for the description of the dilepton 
sources is briefly reviewed. The relevant elementary hadronic reactions 
are systematically discussed. It is demonstrated that 
the resonance model provides an accurate description of 
exclusive vector meson production in nucleon-nucleon 
collisions $NN\rightarrow NN\rho(\omega)$ as well as in pion scattering 
$\pi N\rightarrow N\rho(\omega)$. 
The resonance model allows further to determine the isotopic channels of the 
$N N\rightarrow N N\rho(\omega)$ cross section where no data 
are available. We give iso-spin relations and simple parameterizations 
of the exclusive $N N\rightarrow N N\rho(\omega)$ cross section. 
As discussed in \cite{omega02}, a peculiar role plays thereby the 
$N^*(1535)$ resonance which, fitting available photo-production data, has 
a strong coupling to the $N\omega$ channel. Close to threshold this can lead to 
strong  off-shell contributions to the  $\omega$ production cross section 
\cite{omega02} which are also reflected in the dilepton yields. 
For completeness the dilepton spectra in elementary 
$p+p$ and $p+d$ reactions are reviewed.  

The reaction dynamics of heavy ion collisions is described 
within the QMD transport model \cite{ai91,uma97} which has 
been extended, i.e. the complete set of baryonic 
resonances ($\Delta$ and $N^*$) with masses below 2 GeV has 
been included in the T\"ubingen transport code. A short 
description of the QMD model is given in Sec. IV. 
One purpose of the present investigations is to extract 
information on the in-medium $\rho$- and $\omega$-meson 
widths directly form the BEVALAC data \cite{DLS}. 
The dilepton spectra, distinct from the vector meson masses,
are very sensitive to the vector meson in-medium widths, 
especially the $\omega$-meson. 
The collision broadening is a universal mechanism to increase particle
widths in the medium. E.g., data on the total
photo-absorption cross section on heavy nuclei \cite{fras} provide 
evidence for a broadening of nucleon resonances in a nuclear medium \cite
{Kondratyuk:1994ah}. The same effect should
be reflected in a broadening of the vector mesons in dense matter. 
Since the DLS data show no peak structures which can be attributed 
to the vector meson masses, the problem to extract information 
on possible mass shifts is not yet settled. However, the data 
allow to estimate the order of magnitude of the collision broadening 
of the vector mesons in heavy ion collisions. 

Another  question which is addressed in Sec. III is the role of 
 quantum interference effects. Semi-classical transport models 
like QMD do not keep track of relative phases between amplitudes but 
assume generally that decoherent probabilities can be propagated. On the other 
hand, it has been stressed in several works \cite{zetenyi,lutz03} 
that, e.g., the interference of the isovector-isoscalar 
channels, i.e. the so-called $\rho-\omega$ mixing can significantly alter 
the corresponding dilepton spectra. The $\rho-\omega$ mixing was 
mainly discussed for the dilepton production in $\pi N$ reactions. 
Due to the inclusion of excited mesonic states in the resonance 
decays such interference occurs in our treatment already separately inside each 
isotopic channel. It is natural to 
assume that the interference pattern of the mesonic states will be 
influenced by the presence of surrounding particles. 
In Sect. III, we discuss qualitatively decoherence effects which 
can arise when vector mesons propagate through a hot and dense medium. 
We propose a simple scheme to model this type of decoherence phenomenon
where the environment is treated as a heat bath. This discussion is 
quite general and can be applied, e.g. to the $\rho-\omega$ mixing  
as a special case. It is assumed that before the first 
collision with a nucleon or a pion 
the vector mesons radiate $e^+e^-$ pairs coherently and decoherently afterwards,
since the interactions with a heat bath result in macroscopically 
different final states. As a consequence of charge conservation the 
coherence must be restored in the soft-dilepton limit. The 
present model fulfills this boundary condition. 
The quark counting rules require a destructive interference
between the vector mesons entering into the electromagnetic 
transition form factors of the nucleon resonances. 
Hence, a break up of the coherence results in an increase of 
the dilepton yield below the $\rho$-meson peak. This is 
just the effect observed in the BEVALAC data.
That such a quantum decoherence can at least partially 
resolve the DLS puzzle in heavy ion reactions is demonstrated in Sect.V. 

\section{Elementary sources for dilepton production}

\subsection{Mesonic decays }
At incident energies around 1 GeV meson production (except of the pion) 
is a subthreshold process in the sense that the incident energies lie 
below the corresponding vacuum thresholds. 
The cross sections for meson production ${\cal M}=\eta,\eta^\prime,\rho,\omega,\phi$ 
are small and these mesons, distinct from the pions, 
do not play an essential role for the dynamics of the heavy-ion collisions. 
The production of the mesons 
${\cal M}=\eta,\eta^\prime,\rho,\omega,\phi$ can therefore 
be treated perturbatively. The decays to dilepton pairs take place through 
the emission of a virtual photon. The differential branching ratios for the decay to 
a final state $ X e^{+}e^{-}$ 
\begin{equation}
dB(\mu,M)^{{\cal M}\rightarrow e^{+}e X}
=\frac{d\Gamma (\mu,M)^{{\cal M}\rightarrow e^{+}e^- X}}{\Gamma
_{tot}^{{\cal M}}(\mu)}  \label{BR}
\end{equation}
where $\mu$ is the meson mass and $M$ the dilepton mass are 
taken form \cite{krivo00}. 
These are direct decays ${\cal M}\rightarrow e^{+}e^-$, Dalitz decays 
${\cal M}\rightarrow \gamma e^{+}e^-$, ${\cal M}\rightarrow \pi(\eta) e^{+}e^-$,
and four-body decays ${\cal M}\rightarrow \pi\pi e^{+}e^-$. The experimentally 
known branching ratios are fitted by the 
Vector Meson Dominance (VMD) model and its extension (see below) used in 
 \cite{krivo00}. More exotic decay modes such as, e.g., 
$\phi\to\pi^0e^+e^-,~ \eta\to\pi^+\pi^- e^+e^-$ have recently been 
measured \cite{CMD-2} and are in good agreement with the predictions 
made in \cite{krivo00}. The decay modes determined in 
\cite{krivo00} including channels which contribute to the 
background of the dilepton spectra are taken into account.

\subsection{Resonance decays}

Usually, the description of the decays of baryonic resonances 
$R\rightarrow N\ e^+\ e^-$ is based on the VMD model in its
monopole form, i.e. with only one virtual vector meson ($V = \rho,~\omega$).
As the result, the model provides a consistent description of both, 
radiative $R\rightarrow N\gamma$ 
and mesonic $R\rightarrow NV$ decays. However, a normalization 
to the radiative branchings strongly
underestimates the mesonic ones \cite{resdec,post01,pirner}. 
Possible ways to circumvent this 
inconsistency were proposed in \cite{pirner,post01}. In \cite{pirner} 
a version of the VMD model with vanishing $\rho\gamma$ coupling in the 
limit of real photons ($M^2=0$) was used which allows to fit 
radiative and mesonic decays independently, in \cite{post01} an 
additional direct coupling of the resonances to photons was introduced.  
\begin{figure}[h]
\begin{center}
\leavevmode
\epsfxsize = 12cm
\epsffile[20 430 530 670]{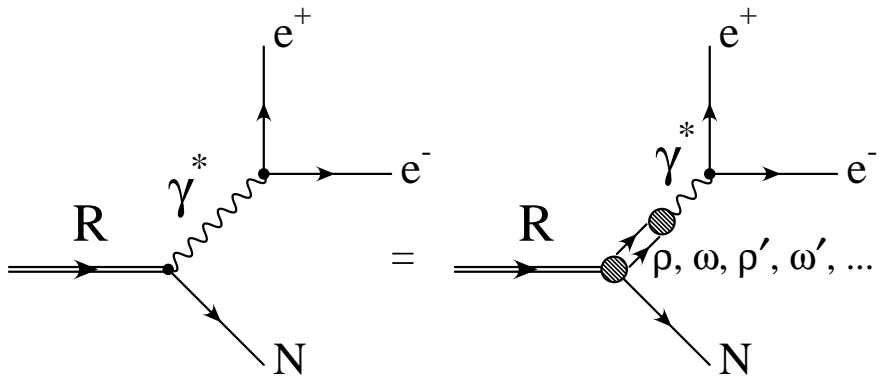}
\end{center}
\caption{Decay of nuclear resonances to dileptons in the extended VMD 
model. The $RN\gamma$ transition form factors contain contributions from 
ground state and excited $\rho$ and $\omega$ mesons.
}
\label{graph1_fig}
\end{figure}
However, apart from that the standard VMD predicts a
$1/t$ asymptotic behavior for the transition form factors. At the 
same time the quark counting rules require a stronger suppression at 
high $t$. A similar problem arises with the $\omega$ Dalitz decay. 
The  $\omega \pi \gamma $ transition form factor shows an 
asymptotic $\sim 1/t^{2}$ behavior \cite{VZ}. It has been 
measured in the time-like region \cite{LGL} and the
data show deviations from the naive one-pole approximation. 
In \cite{krivo00} it was shown that the inclusion of 
higher vector meson resonances in the VMD can resolve this problem 
and provides the correct asymptotics. In \cite{krivo02} the
extended VMD (eVMD) model was used to describe the decay of baryonic 
resonances and in particular to solve the inconsistency between $RNV$ and
$RN\gamma$ decay rates. In the  eVMD model one assumes that 
radial excitations $\rho (1250),~ \rho (1450),\dots$ can 
interfere with the ground state $\rho$-meson in radiative processes. Already in the 
case of the nucleon form factors the standard VMD is not sufficient 
and radially excited vector mesons 
$\rho ^{\prime },$ $\rho ^{\prime \prime }$ $...$ etc. should be
added in order to provide a dipole behavior of the Sachs
form factors and to describe the experimental data \cite{Hohler,Mergell}.
In view of these facts the present extension of the VMD model  is 
more general than the approach pursued in \cite{pirner} since it 
allows not only to describe consistently resonance decays but also 
other observables like the $\omega$ Dalitz decay or the nucleon 
form factor. Here we only briefly sketch the basic 
ideas of the extended vector 
meson dominance (eVMD) model. In Fig. \ref{graph1_fig} the resonance 
decays are schematically displayed for the extended VMD model 
with excited mesons as intermediate states. The interference 
between the different meson families 
plays a crucial role for the behavior of the form factors. Sec. III 
will be devoted to this question. Details of the relativistic calculation 
of the  magnetic, electric, and Coulomb transition form
factors and the branching ratios of the nucleon resonances 
can be found in \cite{krivo02}. 

In terms of the branching 
ratios for the Dalitz decays of the
baryon resonances, the cross section for $e^+ e^-$ production from 
the initial state $X^\prime$ together with the final state 
$N X$ can be written as 
\begin{equation}
\frac{d\sigma (s,M)^{X^{\prime} \rightarrow N X e^{+}e^{-}}}{dM^{2}}
=\sum_{R}\int_{(m_{N}+M)^{2}}^{(\sqrt{s}-m_{X})^{2}}d\mu ^{2}
\frac{d\sigma (s,\mu )^{X^\prime \rightarrow R X}}{d\mu ^{2}}\sum_{V} 
\frac{dB(\mu,M)^{R\rightarrow VN\rightarrow N e^{+}e^{-}}}{dM^{2}}~~.
\label{1}
\end{equation}
Here, $\mu $ is the running mass of the baryon resonance $R$ with the cross
section $d\sigma (s,\mu )^{X^\prime\rightarrow X R}$, 
$dB(\mu ,M)^{R\rightarrow VN\rightarrow N e^{+}e^{-}}$ is the 
differential branching ratio for the
Dalitz decay $R\rightarrow N e^{+}e^{-}$ through the vector meson $V$. Thus 
eq. (\ref{1}) describes baryon induced and pion induced dilepton production, i.e. 
the initial state can be given by two baryons 
$X^\prime = NN,~ NR,~ R^\prime R$ or 
it runs through pion absorption $X^\prime = \pi N$. In the resonance model 
both processes are treated on the same footing by the decay of intermediate 
resonances. 

If the width $\Gamma (R\rightarrow N\gamma ^{*})$ is known, the
factorization prescription \cite{krivo00} can be
used to find the dilepton decay rate
\begin{equation}
d\Gamma (R\rightarrow Ne^{+}e^{-})=\Gamma (R\rightarrow N\gamma
^{*})M\Gamma (\gamma ^{*}\rightarrow e^{+}e^{-})\frac{dM^{2}}{\pi M^{4}},
\label{OK!}
\end{equation}
where 
\begin{equation}
M\Gamma (\gamma ^{*}\rightarrow e^{+}e^{-})=\frac{\alpha }{3}%
(M^{2}+2m_{e}^{2})\sqrt{1-\frac{4m_{e}^{2}}{M^{2}}}  \label{OK!!}
\end{equation}
is the decay width of a virtual photon $\gamma ^{*}$ into the dilepton
pair with the invariant mass $M$.

In the relativistic version of the eVMD model \cite{krivo02} 
which is used here as well as in refs.  \cite{resdec,omega02} the 
decay width $\Gamma (R\rightarrow N\gamma^{*})$ is described by 
three independent transition form factors for resonances with 
spin $J>1/2$ and by only two transition form factors for 
spin-1/2 resonances which follows from the number
of independent helicity amplitudes. In terms of the electric (E), 
magnetic (M), and Coulomb (C) form factors, the decay widths
of nucleon resonances with spin $J = l + 1/2$ into a virtual 
photon with mass $M$ 
has the form \cite{krivo02}

\begin{eqnarray}
\Gamma (N_{(\pm )}^{*} &\rightarrow &N\gamma ^{*})=\frac{9\alpha }{16}\frac{%
(l!)^{2}}{2^{l}(2l+1)!}\frac{m_{\pm }^{2}(m_{\mp }^{2}-M^{2})^{l+1/2}(m_{\pm
}^{2}-M^{2})^{l-1/2}}{\mu^{2l+1}m^{2}_N}  \nonumber \\
&&\left( \frac{l+1}{l}\left| G_{M/E}^{(\pm )}\right| ^{2}+(l+1)(l+2)\left|
G_{E/M}^{(\pm )}\right| ^{2}+\frac{M^{2}}{\mu^{2}}\left| G_{C}^{(\pm
)}\right| ^{2}\right),  
\label{GAMMA_l}
\end{eqnarray}
where $\mu$ refers to the nucleon resonance mass, $m_N$ is the nucleon mass,
$m_{\pm} = \mu \pm m_N$. The signs $\pm$ refer to the natural parity ($1/2^-,
3/2^+, 5/2^-,$ ...) and abnormal parity ($1/2^+,
3/2^-, 5/2^+,$ ...) resonances. $G_{M/E}^{\pm}$ means $G_{M}^{+}$ or $G_{E}^{-}$. 
The above equation is valid
for $l>0$. For $l=0$ ($J=1/2$), one gets 
\begin{eqnarray}
\Gamma (N_{(\pm )}^{*} &\rightarrow &N\gamma ^{*})=\frac{\alpha }{8\mu}%
(m_{\pm }^{2}-M^{2})^{3/2}(m_{\mp }^{2}-M^{2})^{1/2}  \nonumber \\
&&\left( 2\left| G_{E/M}^{(\pm )}\right| ^{2}+\frac{M^{2}}{\mu^{2}}\left|
G_{C}^{(\pm )}\right| ^{2}\right).  \label{GAMMA_0}
\end{eqnarray}
In \cite{krivo02} the extended VMD model was applied in 
a fully covariant form to the 
description of the transition form factors of the nucleon
resonances with arbitrary spin and parity. The decay widths are then 
given in terms of covariant amplitudes which can be converted to 
magnetic, electric and Coulomb transition form factors. To constrain 
the asymptotics quark counting rules were used. 
The free parameters of the model are fixed by fitting
the experimental data on the photo- and electro-production amplitudes and by
fitting the results of multichannel $\pi N$-scattering partial-wave
analysis and quark model predictions for these amplitudes. 
In the relativistic treatment the 
number of intermediate $\rho$ (or $\omega$) 
states which have to be taken into account to describe the  
magnetic, electric and Coulomb transition form factors depends on 
the resonance spin $J$, i.e. $J-\frac{1}{2}+3$ mesons have to be included 
in the minimal version of the eVMD model. Since we consider resonances 
with spins ranging from $\frac{1}{2}$ up to $\frac{7}{2}$ the number of 
$\rho$ states is maximally 6. The following masses have been 
used: 0.769, 1.250, 1.450, 1.720, 2.150, 2.350 (in GeV). Within this 
description dilepton branching ratios 
were determined quantitatively for baryonic resonances with masses 
below 2 GeV. In particular, a simultaneous 
description of radiative and mesonic decays could be achieved. 
For further details we refer the reader to ref. \cite{krivo02}.

\subsection{Vector meson production in $NN$ collisions}
Cross sections for the direct vector meson production ($V=\rho,~\omega,~\phi$) 
in nucleon-nucleon collisions $\sigma^{NN\rightarrow XV}$ can 
e.g. be taken from \cite{sibirtsev96,sibirtsev97}. These are 
parameterizations of the inclusive production cross sections 
in proton-proton reactions ($pp\rightarrow XV$) fitted to 
experimental data in combination with 
LUND string model predictions \cite{sibirtsev97} and exclusive 
cross sections determined in a one-pion-exchange picture \cite{sibirtsev96}.  
However, in heavy ion reactions at subthreshold energies, i.e. 
in the BEVALAC and SIS domain, one can expect that significant 
strength of the dilepton yield originates from the decay of vector 
mesons, in particular the $\rho$, which are far off-shell with masses 
well below their pole values. Such processes give contributions to 
the cross sections below the sharp threshold $\sqrt{s_0}=2m_N +m_V$ 
with $m_V$ the pole mass. Subthreshold meson production can be naturally described 
through the decay of baryonic resonances 
\cite{resdec,post01,titov,zetenyi,BCM,resgiessen2}. Around threshold the final 
states consist only of two nucleons and the corresponding meson. 
These are the processes which are relevant in heavy ion reactions 
at intermediate energies in the BEVALAC and GSI range, i.e. at bombarding 
energies below 2 AGeV. Due 
to the moderate incident energies involved in the elementary reactions 
it is sufficient to consider exclusive meson production.  
Since the production of vector mesons through 
the decay of baryonic resonances gives a significant contribution 
to the total cross section one has thereby to avoid the problem of 
double counting between the dilepton production via baryonic resonances 
and those originating from other sources. A detailed discussion 
of the double counting problem in  nucleon-nucleon collisions can be found in 
\cite{resdec}. 

The vector meson production cross section is now given as follows
\begin{equation}
\frac{d\sigma (s,M)^{NN\rightarrow NNV}}{dM^{2}}
=\sum_{R}\int_{(m_{N}+M)^{2}}^{(\sqrt{s}-m_{N})^{2}}d\mu ^{2}\frac{ 
d\sigma (s,\mu )^{NN\rightarrow NR}}{d\mu ^{2}}
\frac{dB(\mu,M)^{R\rightarrow VN}}{d M^{2}}~~.
\label{sigNNV}
\end{equation}
The cross sections for the resonance production are given by 
\begin{equation}
d\sigma (s,\mu )^{NN\rightarrow NR} = 
\frac{|{\cal M}_R|^2 ~ p_f}{16 p_i s\pi}~dW_R(\mu)
\label{sigNR}
\end{equation}
with the final c.m. momentum 
\begin{eqnarray}
p_f = p^*(\sqrt{s},\mu,m_{N})
= \frac{\sqrt{(s-(\mu+m_N)^2)(s-(\mu-m_N)^2)}}{2\sqrt{s}}
\label{cr4}
\end{eqnarray}
and the initial c.m. momentum $p_i$. The mass distributions 
$dW_R(\mu)$ of the resonances are usual Breit-Wigner distributions 
\begin{equation}
dW_R(\mu) = 
\frac{1}{\pi} \frac{\mu \Gamma^R (\mu) d\mu^2 }
{(\mu^2 - m_{R}^2)^2 +(\mu\Gamma_{\rm tot}^R(\mu))^2}
\label{BW}
\end{equation}
where $\mu$ and $m_R$ are the running and pole masses, respectively, 
and $\Gamma(\mu)$ is the mass dependent resonance width. The 
matrix elements ${\cal M}_R$ are taken from \cite{Teis,Bass} where 
they have been adjusted to one and two-pion production data. 
For the description of the $\rho$ and $\omega$ production in $NN$ and 
$\pi N$ reactions we consider the same set of 
resonances which has been used in refs. \cite{resdec,omega02}. It 
includes only the well established ($4*$) resonances listed by 
the PDG \cite{pdg} and is smaller than the complete set of 
resonances included in the QMD model. This set of resonances is, 
however, sufficient to describe the $NN$ and $\pi N$ vector meson production data. 
The corresponding decay widthes $\Gamma_{N\rho}$, $\Gamma_{N\omega}$ 
at the resonance pole masses are given in Tables 
\ref{nstar_tab}, \ref{delta_tab}. Off-shell the normalization of the 
total widths is ensured by the same procedure as used in ref. \cite{omega02}. 
\begin{figure}[h]
\begin{center}
\leavevmode
\epsfxsize = 14cm
\epsffile[20 50 550 390]{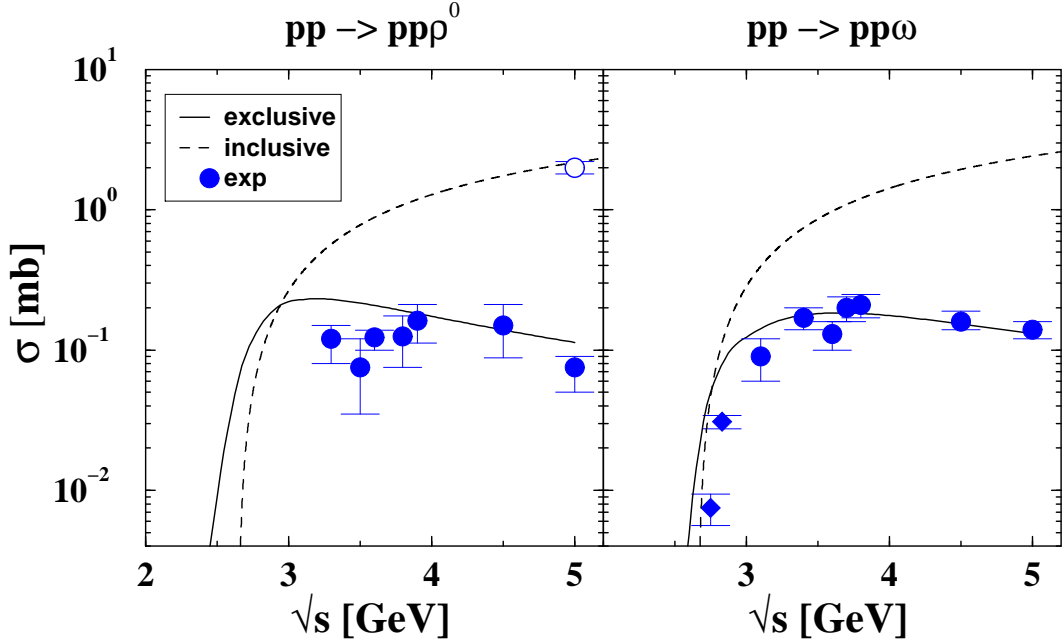}
\end{center}
\caption{Cross sections for the   $\rho^0$ and $\omega$ production 
in proton-proton reactions. The exclusive vector meson 
cross sections through the decay of baryonic resonances are compared 
to data and to the inclusive cross sections of 
\protect\cite{sibirtsev97}. For the $\rho^0$ also one data point 
(open circle) for the inclusive cross section is shown.  
The $\omega$ data are taken from \protect\cite{cosy01} (diamonds) 
and \protect\cite{disto01,flaminio} (circles).
}
\label{sig1_fig}
\end{figure}
In Fig. \ref{sig1_fig} the resonance contributions 
$pp\rightarrow pR\rightarrow pp \rho^0 (\omega )$ to the exclusive 
$\rho^0$ and $\omega$ production are compared to the inclusive 
cross section from \cite{sibirtsev97} and to corresponding experimental 
data for the exclusive cross sections. It can be seen from there that the 
exclusive $pp\rightarrow pp \rho^0 (\omega) $ cross sections can be saturated 
by the excitation of intermediate resonances. 
In the present calculations the dilepton production via the decay of baryonic 
resonances (\ref{1}) runs over intermediate vector mesons 
with mass $M$ which can be off-shell. Therefore, in eq. (\ref{sigNNV}) 
the thresholds for the production of a vector meson with mass $M$ 
are given by the two pion threshold $2M_N + 2m_\pi$ for the $\rho$, respectively 
the three pion threshold $2M_N + 3m_\pi$ for the $\omega$. 
This is in contrast to parameterizations of the elementary cross 
sections \cite{sibirtsev96,sibirtsev97} where 
vector mesons are produced with sharp thresholds given by their 
pole masses ($\sqrt{s_0}=2M_N + m_V$). 

The subthreshold production 
of vector mesons results in a significant strength near 
$\sqrt{s_0}$ and below. Due to the 
broad  $\rho$ width this gives the dominant contribution 
to the total cross section around threshold and explains 
the differences between our calculation and the parameterization 
of \cite{sibirtsev97}. The subthreshold production 
is of course smaller for the $\omega$. However, as discussed e.g. 
in \cite{hibou99} at threshold also in the case of the $\omega$ 
a large amount of the 
cross section can originate from subthreshold $\omega$ production. 
On the other hand, the inclusion of subthreshold meson production 
makes the comparison with data more difficult since the experimental 
identification by correlated pions misses strength from such subthreshold 
processes \cite{hibou99}. Consequently, two recent data points from 
the COSY-TOF Collaboration \cite{cosy01} 
for $pp\rightarrow pp \omega$ are overestimated in Fig. \ref{sig1_fig}.   
However, in $pp$ reactions at low incident energies the subthreshold contribution 
dominates the dilepton yield in the mass region between the $\eta$ 
and the $\rho-\omega$ peak \cite{resdec}. 

\begin{figure}[h]
\begin{center}
\leavevmode
\epsfxsize = 14cm
\epsffile[20 50 550 390]{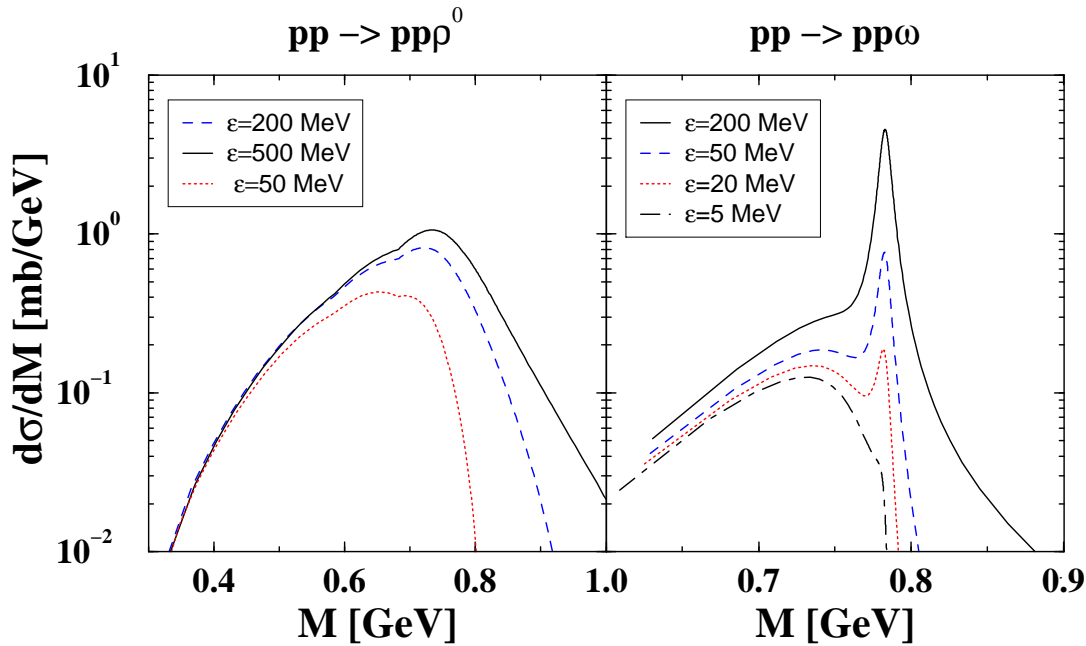}
\end{center}
\caption{Differential cross sections $d\sigma/dM$ 
for the $\rho^0$ and $\omega$ production 
in proton-proton reactions as a function of the meson mass $M$. 
The cross sections are shown for various values of the excess 
energy $\epsilon = \protect\sqrt{s}-(2m_N+m_V)$ where $m_V$ is given 
by the $\rho$ and $\omega$ pole masses.}
\label{sig2_fig}
\end{figure}

The importance of the subthreshold contributions where the $\rho$ 
and $\omega$ are produced with masses far below their pole values 
can be estimated from Fig.\ref{sig2_fig}. Here differential cross sections 
$d\sigma/dM$ are shown as functions of the meson mass $M$ 
for the same reactions as in Fig.\ref{sig1_fig}. The cross sections are 
calculated at different energies, translated into the excess energy 
$\epsilon = \sqrt{s} -  \sqrt{s_0}$. It is clear that close to 
``threshold'' the cross sections are dominated by ``subthreshold'' 
production where the vector mesons are produced off-shell. The physical 
thresholds are given by $2m_\pi$ for the $\rho$ and $3m_\pi$ 
for the $\omega$, respectively. Experimentally these off-shell 
contributions can hardly be distinguished from the  general pionic background 
in coincidence measurements and are generally treated as background. 
Due to the large $\rho$ width it is nearly impossible to distinguish the 
 $\rho$ peak from this background contribution which makes it 
impossible to identify the  $\rho$ experimentally at small excess
energies.

\begin{figure}[h]
\begin{center}
\leavevmode
\epsfxsize = 10cm
\epsffile[60 50 430 410]{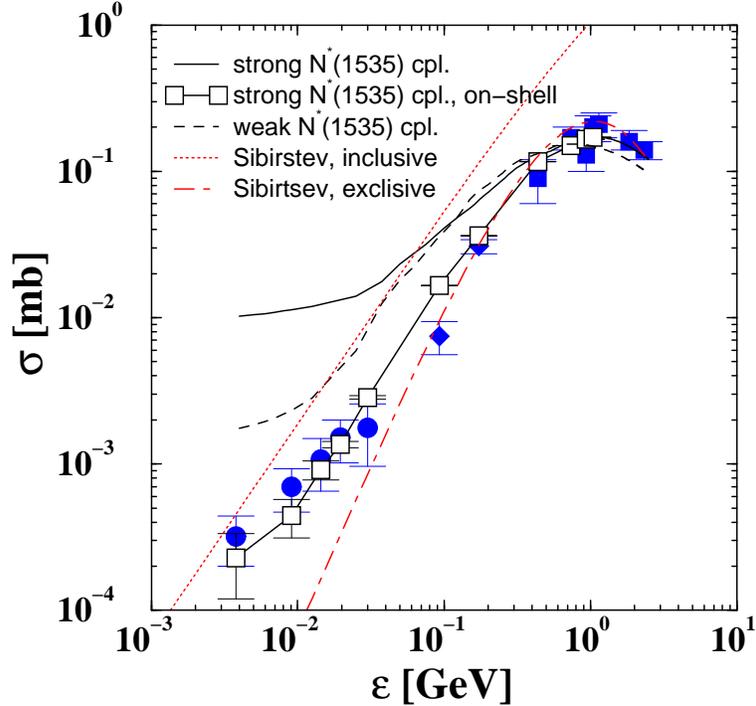}
\end{center}
\caption{Exclusive $pp\rightarrow pp\omega$ cross section 
obtained in the resonance model as a function of the excess energy 
$\epsilon$. The solid curve shows the full cross 
section (strong $N^*(1535)N\omega$ coupling) including off-shell 
contributions while the squares show the experimentally 
detectable on-shell part of the cross section. The dashed 
curves show the corresponding cross section obtained with weak 
$N^*(1535)N\omega$ coupling. The dotted curve is a 
parameterization of the inclusive cross section from 
\protect\cite{sibirtsev97}. 
Data are taken from \protect\cite{hibou99,cosy01} and 
\protect\cite{disto01,flaminio}. 
}
\label{sigom_fig}
\end{figure}

The situation is more complicated for the $\omega$. A detailed 
investigation of the $\omega$ production in $pp$ reactions within the 
framework of the resonance model was performed in \cite{omega02}.
Among the considered resonances the $N^*(1535)$ 
turned out to play a special role for the $\omega$ production. The 
reason is a large decay mode of this resonance 
to the $N\omega$ channel in a kinematical 
regime where the $\omega$ is far off-shell. A strong
$N^*(1535)N\omega$ coupling is implied 
by the available electro- and photoproduction data \cite{krivo02}. 
As a consequence large off-shell contributions in the $\omega$ production 
cross section appear. In particular close to threshold the 
off-shell production is dominant \cite{omega02}. 
This part of the cross section can, however, experimentally not be 
identified and is currently attributed 
to the experimental background. To compare to data we applied in \cite{omega02} 
the same procedure as experimentalists: The theoretical 
"background" from the off-shell 
production was subtracted and only the measurable pole part 
of the cross section was taken into account. Doing so, without 
adjusting any new parameters the available data are 
accurately reproduced from energies very 
close to threshold \cite{hibou99,cosy01} up to energies 
significantly above threshold \cite{disto01,flaminio}. 
At small excess energies the full 
cross section shown in Fig.\ref{sigom_fig} is about one order of magnitude 
larger than the measurable pole part. 
Since the $\omega$ cross section depends crucially on the 
role of the $N^*(1535)$ in \cite{omega02} 
we considered also an alternative possible scenario: 
The $N\omega$ decay of the $N^*(1535)$ resonance has not directly been
 measured and the existing $N\rho$ data 
leave some freedom to fix the eVMD model parameters. A different 
normalization to the $N\rho$ channel, making thereby use of an 
alternative set of quark model predictions, allows to reduce the 
$N\omega$ decay mode by maximally a factor of 6 to 8, however, at 
the expense of a slightly worse reproduction of the existing data set. 
With the reduced $N\omega$ coupling the off-shell contributions 
are substantially reduced. However, the pole part of the cross 
section leads to a significant overestimation of the 
experimental data around and several 100 MeV above threshold. 
The $\rho$ production turned out to be practically independent on 
the choice of the two different parameter sets. 
In \cite{omega02} we concluded that based on the  $pp\omega$ data 
it will be not possible to decide whether the  $\omega$ production 
is accompanied by strong off-shell contributions close to 
threshold or not, because this part of the cross section is experimentally 
not accessible. However, these off-shell contributions 
fully contribute to the dilepton yield from $\omega$ decays. 
Therefore, in Sec.V we consider two different scenarios for 
the dilepton production through $\omega$ decays:
\begin{enumerate}
\item $\omega$ production through baryonic resonances with strong 
 $N^*(1535)\omega$ coupling, leading to large off-shell contributions 
around threshold.

\item $\omega$ production through baryonic resonances with weak 
 $N^*(1535)\omega$ coupling, leading to small off-shell contributions 
around threshold.
\end{enumerate}

\begin{figure}[h]
\begin{center}
\leavevmode
\epsfxsize = 14cm
\epsffile[20 50 550 390]{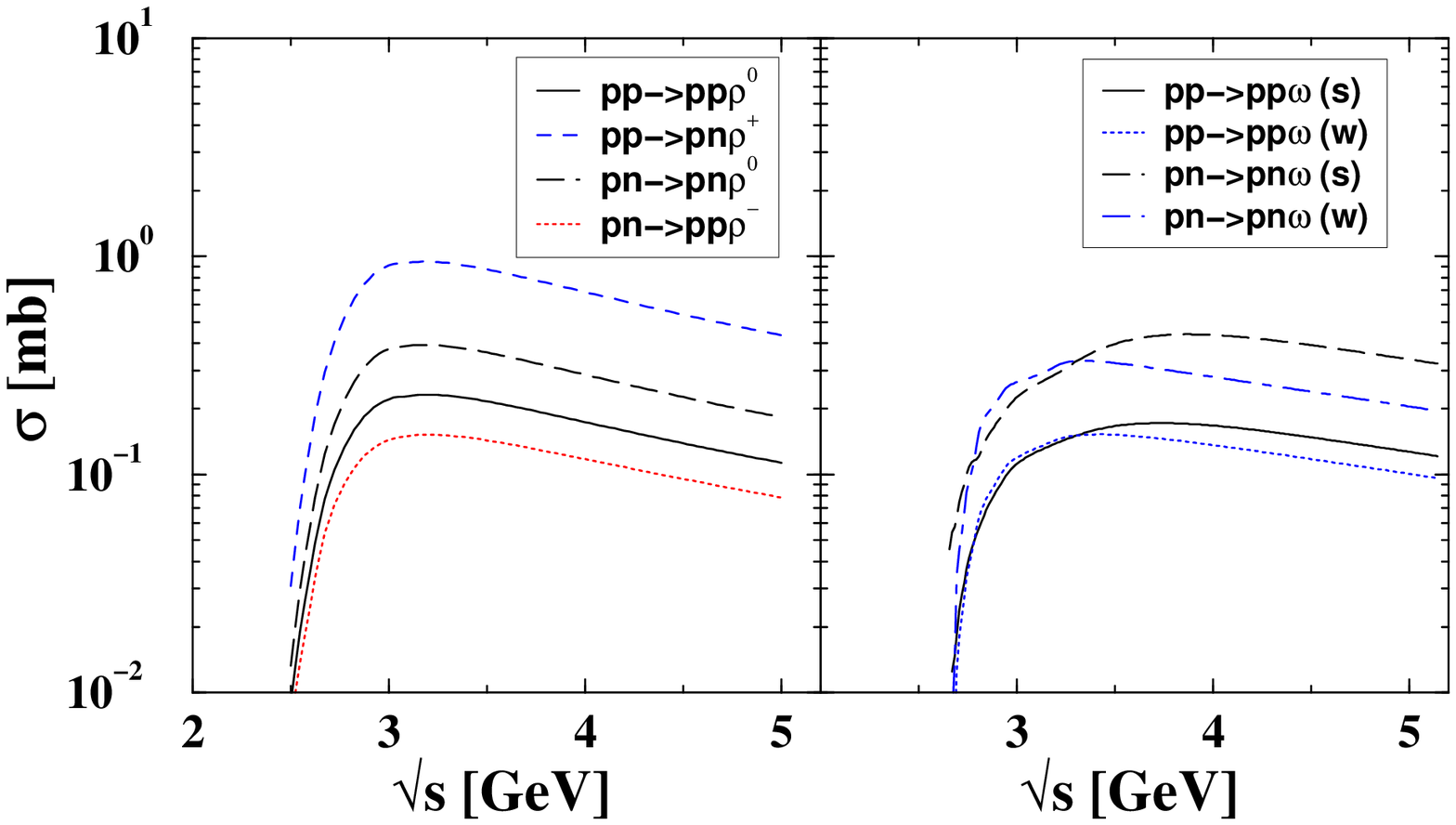}
\end{center}
\caption{Left: isospin dependence of the exclusive $ NN\rightarrow NN\rho$ 
cross section assuming isospin independent matrix elements for the 
resonance production.\\
Right: isospin dependence of the exclusive $ NN\rightarrow NN\omega$ 
cross section. The isospin dependence of the $N^*(1535)$ is taken into 
account. We distinguish between a strong (s) and a weak (w) $N^*(1535)N\omega$ 
coupling.
}
\label{iso_fig}
\end{figure}

Fig.\ref{sigom_fig} summarizes the different possibilities to treat the 
$\omega$ production in elementary $NN$ reactions. The different cross 
sections are shown as functions of the excess energy $\epsilon$. The 
resonance model, assuming a large $N^*(1535)N\omega$ coupling, leads to 
very accurate description of the measured on-shell cross section. 
It has, however, a very strong off-shell component which fully 
contributes to the dilepton production. The weak coupling scenario, on 
the other side, has only small off-shell component but the 
reproduction of the data is relatively poor in the low energy regime. 
The parameterization of the inclusive cross section 
$\sigma^{pp\rightarrow\omega X}=2.5(s/s_0-1)^{1.47}(s/s_0)^{-1.11}$ 
\cite{sibirtsev97} which has been used in \cite{BCRW98,BCEM} is also 
shown for comparison. 
If cross sections are based on fits to data iso-spin factors are 
usually obtained from the corresponding Clebsh-Gordon coefficients under 
the assumption of totally iso-spin independent matrix elements. Such an assumption 
is, however, crude. It is not possible to fix the two different 
iso-spin amplitudes of the $\rho NN$ final state and their relative 
phases solely from measured cross sections and without further model 
assumptions. In the resonance model the iso-spin dependence 
of the cross sections is well defined by coupling the final states 
to $N\otimes [N\otimes\rho ]$. In the $N\rho$ system the $I=\frac{3}{2}$ 
amplitude contains all $\Delta$-resonances whereas the $I=\frac{1}{2}$ 
contains the contributions form the $N^*$s. Since the resonance amplitudes 
are summed incoherently the cross section can be easily be decomposed 
into the corresponding iso-spin contributions. The isotopic channels of the 
$NN\rightarrow NN\rho$ cross section are then uniquely fixed by 
\begin{equation}
\sigma (NN\rightarrow NN\rho ) = \alpha \sigma_{\frac{3}{2}} 
+ \beta  \sigma_{\frac{1}{2}}
\label{para1}
\end{equation}
where $\alpha,~\beta$ are determined from the corresponding 
Clebsh-Gordon coefficients. The coefficients are summarized in 
Table \ref{iso_tab1}. Fig. \ref{iso_fig} shows the corresponding 
contributions $\sigma_{\frac{3}{2}}$ and $\sigma_{\frac{1}{2}}$ 
originating from the sum over $\Delta$ and $N^*$ resonances, 
respectively, and the different isospin channels of the 
$NN\rightarrow NN\rho$ cross section. The isospin dependence is 
significant. The $pn\rightarrow pn\rho^0$ channel is about two times 
and the $pn\rightarrow pn\rho^+$ about four times larger than the 
measured $pp\rightarrow pp\rho^0$ channel.

The two isotopic channels $\sigma_{\frac{3}{2}}$ and $\sigma_{\frac{1}{2}}$ 
can be parameterized in the form
\begin{equation}
\sigma_{\frac{3}{2},\frac{1}{2}} = \frac{ a_{1} (\sqrt{s}- a_{2})^{a_{3}} }
{(\sqrt{s}- a_{4})^2 + a_{5} }
\label{sigpara1}
\end{equation}
with the coefficients $a_1 = 0.7813(0.334)~,a_2=2.512(2.508),~
a_3= 1.206(1.135),~a_4=2.736(2.426),~a_5 = 0.293(0.412)$ for the 
$I=\frac{3}{2}(\frac{1}{2})$ channels. 
A parameterization of $\sigma (pp\rightarrow pp\omega)$ by (\ref{sigpara1}) 
yields the following coefficient: $a_1= 0.4921,~a_2 = 2.656,~a_3 = 0.7529,~
a_3 = 2.6812,~a5 = 1.8395$. Note that the thresholds for the 
parameterizations (\ref{sigpara1}) are given by the $a_2$ values and 
account only partially for the subthreshold contributions in the cross 
sections.
The isotopic relations given in Tables I and II  are derived under the 
assumption of isospin independent matrix elements ${\cal M}_R$ for the 
resonance production (\ref{sigNR}). This assumption is justified for 
all resonances except of the $N^*(1535)$ \cite{Teis}. For this resonance the
$pn\rightarrow pn^*(n p^*)$ cross section is 
known to be about 5 times larger than 
for  $pp\rightarrow pp^*$ \cite{Teis}. This fact is also reflected in the isotopic 
relation for the $\eta$ production to which the $N^*(1535)$ has a large 
branching ratio. If we take the enhancement of the $N^*(1535)$ matrix 
element in the $pn$ channel 
by a factor 5 into account, the $pn\rightarrow pn \rho^0$ cross section 
shown in Fig. \ref{iso_fig} is shifted upwards by 10\% and the 
$pn\rightarrow pp \rho^-$ cross section by 20\%.

For the $\omega$ production only $N^*$ resonances contribute and thus 
the naive isospin relation would imply $
\sigma (pn \rightarrow pn\omega) = \sigma (pp\rightarrow pp\omega)$. 
However, in this case  the strongly isospin dependent $N^*(1535)$ 
production cross section has a large influence which depends of 
course on the strength of the $N^*(1535)N\omega$ coupling. In the 
case of a weak coupling the $pn\rightarrow pn\omega$ channel is 
enhanced by a factor two, in the case of a strong coupling even by 
a factor of three. For all other resonances which contribute to 
the $ NN\rightarrow NN\omega$ cross section shown in  Fig. \ref{iso_fig} 
(right) isospin symmetric matrix elements are assumed.

\subsection{Vector meson production in $\pi N$ collisions}
\begin{figure}[h]
\begin{center}
\leavevmode
\epsfxsize = 14cm
\epsffile[20 50 550 390]{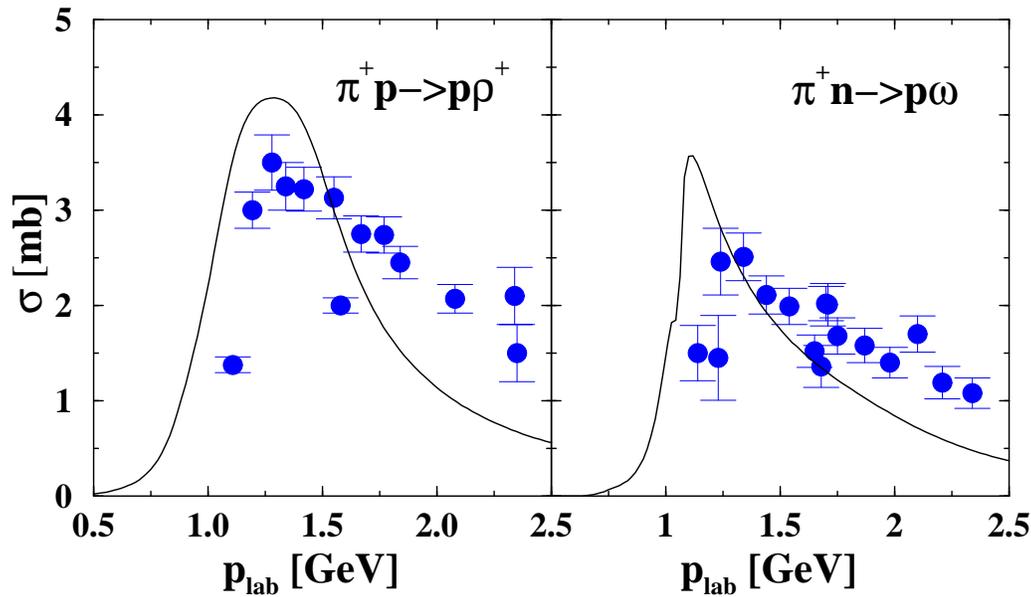}
\end{center}
\caption{Exclusive $\pi^+ p\rightarrow p\rho^+$ and 
$\pi^+ n\rightarrow p\omega$ cross sections obtained within the 
resonance model. The experimental $\pi^+ p\rightarrow p\rho^+$ are taken from 
\protect\cite{pirho_dat}.  
}
\label{piV_fig1}
\end{figure}

Similar as in the previous case the pion induced vector meson production 
can be parameterized and fitted to existing data. E.g. in \cite{sibirtsev97} 
the exclusive and inclusive $\pi N\rightarrow N\rho (\omega,~\phi)$ 
cross sections have been fitted to data and LUND string model 
predictions. In the present work we describe the exclusive 
cross sections again microscopically within the resonances model 
\begin{eqnarray}
\frac{d\sigma (s,M)^{\pi N\rightarrow NV}}{dM^{2}}
&=&\sum_{R}d\sigma (s,\mu )^{\pi N\rightarrow R}\frac{dB(\mu
,M)^{R\rightarrow VN}}{d M^{2}} 
\nonumber \\
&=& \sum_{R}\frac{(2j_R+1)}{(2j_N+1)} \frac{\pi^2}{p_{i}^2} 
\Gamma^{R}_{N\pi} (\mu) dW_{R} (\mu)
\frac{dB(\mu,M)^{R\rightarrow VN}}{d M^{2}}
\label{sigpiNV}
\end{eqnarray}
where $j_R$ is the resonance spin, $j_N$ the nucleon spin and $p_i$ the 
$\pi N$ c.m. momentum. As in the previously discussed $NN$ reactions 
the cross sections are calculated as an incoherent sum over all 
resonances. The same approximation has also been used in other works 
\cite{post01}. Fig. \ref{piV_fig1} shows the corresponding  
$\pi^+ p\rightarrow p\rho^+$ and 
$\pi^+ n\rightarrow p\omega$ cross sections. At laboratory momenta 
below 1.5 GeV the existing data are generally well 
reproduced. Close to threshold the same phenomenon as in the $NN$ 
reactions occurs, i.e. the off-shell meson production gives a large 
contribution to the total cross section. Again low energy data which 
exist in the case of the $\omega$ are overpredicted by the  calculations. 
At higher energies the agreement with experiment is very reasonable, 
both for the $\rho$ and the $\omega$. However, at momenta above $1.5\div 2$ GeV 
the data are generally underpredicted. 
As can be seen form Fig. \ref{pion_fig}
also the total $\pi^+ p\rightarrow X$ and  $\pi^- p\rightarrow X $ 
cross sections can only be well 
described up to pion laboratory momenta around 1.2-1.5 GeV. For the 
determination of the inclusive pion cross sections all baryonic 
resonances given in Tables \ref{nstar_tab}, \ref{delta_tab} are taken into account.  
Nevertheless, at large $p_{\rm lab}$ the contributions of even 
higher lying resonances or 
other direct processes seem to be missing. In the determination of the 
vector meson production cross sections we rely on the same set of 
resonances which has been used for $NN$ reactions dicussed in the 
previous subsection. 
Thus some of the higher lying and insecure resonances included 
in Fig. \ref{pion_fig} are not taken into account here. A substantial 
missing strength in the 
$\pi N \rightarrow N\omega$ cross section at large values of $p_{\rm lab}$ 
has also been found in \cite{post01}. Compared to \cite{post01} our 
results for the cross sections are generally somewhat larger and thus 
in better agreement with the data. The reason lies in a different 
determination of the resonance decay modes to vector mesons 
within the extended vector dominance model \cite{resdec}.

As in the case for the $NN$ reaction iso-spin relations are determined 
by the composition into contributions from $\Delta$ and $N^*$ resonances. 
Using the same representation as in Eq. (\ref{para1}), 
$\sigma (\pi N\rightarrow N\rho ) = \alpha \sigma_{\frac{3}{2}} 
+ \beta  \sigma_{\frac{1}{2}}$ the corresponding iso-spin coefficients 
are given in Table \ref{iso_tab2}.  

In summary, at high energies one has to restrict oneself to 
phenomenological fits to data 
\cite{sibirtsev97} or include string model excitations. For the 
SIS energy domain where vector mesons are predominantly produced 
subthreshold the present model gives a reliable description of 
the vector meson production in $\pi N$ reactions. 

\subsection{Dilepton production in in $pp$ and $pd$ reactions}
\begin{figure}[h]
\begin{center}
\leavevmode
\epsfxsize = 10cm
\epsffile[25 45 510 760]{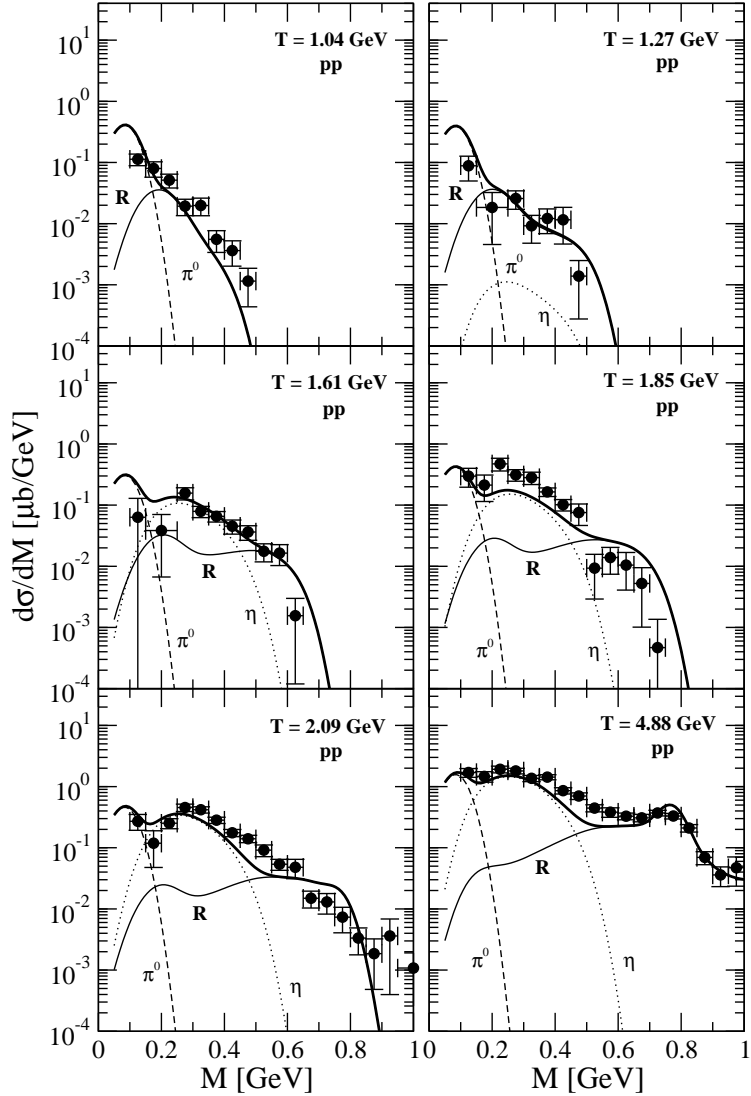}
\end{center}
\caption{The differential $pp\rightarrow e^+e- X$ cross sections 
at various proton kinetic energies are compared to the DLS data 
\protect\cite{DLS2}. 
}
\label{DLS_fig1}
\end{figure}

Before turning to heavy ion collisions we will consider the dilepton 
production in elementary reactions. Dilepton spectra  in 
proton-proton and proton deuteron reactions have been measured by 
the DLS Collaboration in the energy range from $T=1\div 5$ GeV \cite{DLS2}. 
The application of the present model to the dilepton production 
in $pp$ reactions has in detail been discussed in \cite{resdec}. 
For completeness we show the corresponding results and the 
comparison to the DLS data \cite{DLS2} in Fig. \ref{DLS_fig1}. 
The agreement with the available data is generally reasonable, 
i.e. of similar quality as obtained in previous calculations by 
Ernst et al. \cite{ernst} and Bratkovskaja et al. \cite{BCM}. 
As in \cite{ernst} we observe a slight underestimation of 
the experimental dilepton yield at the two highest energies $T$=2.09 and 
4.88 GeV in the mass region below the $\rho -\omega$ peak. Here the 
knowledge of the inclusive cross section with multi-pion final 
channels starts to play an important role. In \cite{resdec} the 
multi-pion production was estimated within an semi-empirical model 
which is slightly modified in the present case. However, results 
are very similar to our previous calculations \cite{resdec}. 
It should be noted that the dilepton yields in $pp$ reactions were 
obtained with the strong $N^*(1535)-N\omega$ decay mode. As briefly 
sketched in Sec. II and in detail discussed in \cite{omega02} the 
strong coupling mode is the result of the eVMD fit to the 
available photo- and meson-production data \cite{krivo02}. 
It leads to sizable contributions from off-shell $\omega$ production around threshold 
energies which are, however, experimentally not accessible in 
$pp\rightarrow pp\omega$ measurements. On the other side, these off-shell 
$\omega$'s fully contribute to the dilepton yield. The off-shell 
contributions lead generally to an enhancement of the dilepton 
yield in the mass region below the $\omega$ peak, in particular 
at incident energies where the $\omega$ is dominantly produced 
subthreshold. In contrast to \cite{ernst,BCM} where the $\omega$ is 
treated as an elementary particle (with fixed mass $m_\omega$=782 MeV) 
in our approach the off-shell $\omega$ production starts 
at the three-pion threshold. Thus subthreshold 
$\omega$ production appears already in elementary reactions. 
As can be seen from Fig.\ref{DLS_fig1} the scenario of large 
off-shell $\omega$ contributions which are the consequence of the 
strong $N^*(1535)-N\omega$ coupling are consistent with the 
experimental $pp$ dilepton yields in the energy range of $T=1.04\div 1.61$ 
GeV. At higher energies this off-shell production becomes 
negligible \cite{omega02}. 
\begin{figure}[h]
\begin{center}
\leavevmode
\epsfxsize = 10cm
\epsffile[25 45 510 760]{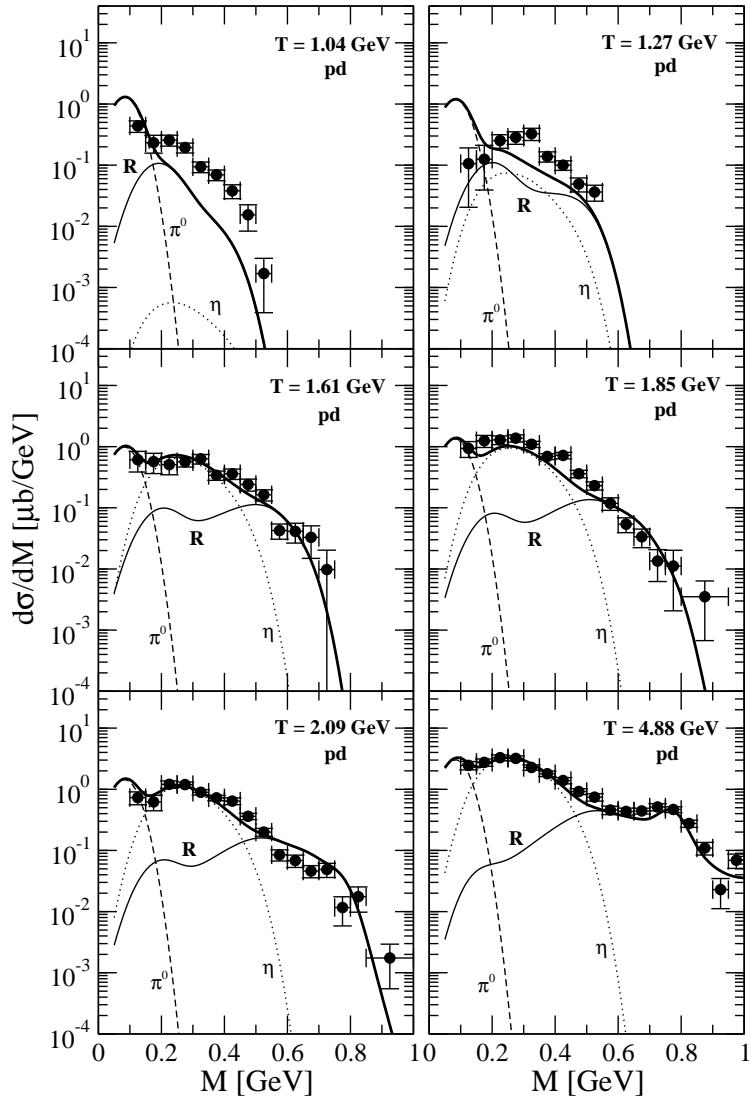}
\end{center}
\caption{The differential $pd\rightarrow e^+e^- X$ cross sections
at various proton kinetic energies are compared to the DLS data
\protect\cite{DLS2}.
}
\label{DLS_fig2}
\end{figure}
The situation becomes more complicated when proton-deuteron 
reactions are considered. Compared to the $pp$ case one has here 
two important modifications: First the Fermi motion of the 
proton and neutron constituents inside the 
deuteron and secondly, the isotopic relations between
the $p p$ and $p n$ contributions to the dilepton production.
Only few isotopic relations for the meson production are 
experimentally fixed. Most isospin relations have to be derived 
from model assumptions (see also Sec. II). For the dilepton
production in $p N$ collision we distinguish generally
between three different channels
$$p N \rightarrow N R \rightarrow NN \pi^0~;~\pi^0\rightarrow \gamma e^+ e^-
~,$$
$$p N \rightarrow N R \rightarrow NN \eta ~;~ \eta\rightarrow \gamma e^+ e^-~
 ,$$
$$p N \rightarrow N R \rightarrow NN  e^+ e^-~,$$
where $R$ is either a nucleon resonance $N^*$ or a $\Delta$
resonance. The last channel contains all contributions which run
over intermediate $\rho$ and $\omega$ mesons.
For the first channel we use here the following isotopic relation for
fixed two-nucleon final states ($NN$):
$p p:p n =1:1 (5)$ if the intermediate resonance is $R=N^*(N^*(1535))$
and $p p:p n =1:2$ for $R=\Delta$ \cite{Teis}.
To the $\eta$ production only the $N^*(1535)$
contributes \cite{Teis} and thus the isotopic relation is $p p:p n =1 : 5$.
The third channel has the same isotopic relations as the first channel
if one assumes that intermediate $\rho$ and $\omega$ mesons are
effectively not interfering in $p n$ collision. The latter means that
for two equally probable
reactions $p n \rightarrow p R^0$ and $p n \rightarrow n R^+$
the radiative decays of $R^0$ and $R^+$ resonances have no
$\rho -\omega$ interference when summed. Then the isospin relations for
the $\rho^0$ and $\omega$ can be read from Table I.
The Fermi motion of the constituents inside
the deuteron is taken into account using the experimental momentum
distribution of the bound proton which was
obtained by electron scattering \cite{fmopid}.

\begin{figure}[h]
\begin{center}
\leavevmode
\epsfxsize = 10cm
\epsffile[40 60 680 550]{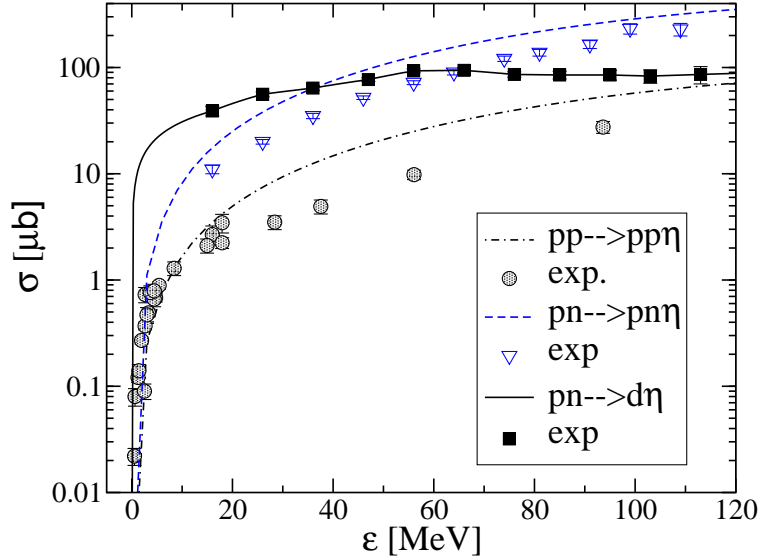}
\end{center}
\caption{Experimental data on near to the threshold cross sections of
the reactions: $pp\rightarrow pp\eta$ (circles) 
\protect\cite{hibou98,cosy11,Calen1,Chia,berg},
$pn\rightarrow pn\eta$ (triangles) and $pn\rightarrow d\eta$ (squares) were taken from
\protect\cite{Calen}. The curves show the corresponding model cross sections.
}
\label{eta_fig}
\end{figure}

At the two lowest incident proton energies of $T=1.04$ GeV and $T=1.27$ GeV 
the threshold effects for the $\eta$ production become extremely important.
For a target nucleon at rest the $\eta$ production is far below threshold
at $T=1.04$ GeV ($\epsilon =-84$ MeV, $\epsilon$ is the excess energy
in the center of mass system) and slightly above
threshold at $T=1.27$ GeV ($\epsilon=6.4$ MeV). The Fermi motion 
of the proton and neutron constituents inside the
deuteron increases the accessible $\epsilon$ values. In the 
present calculations experimental results form 
 electron scattering \cite{fmopid} are used to model the 
proton and neutron momentum distributions. It is further known 
from experiment \cite{Calen} that close to threshold the
$pn\rightarrow d\eta$ cross section is much larger (by a factor $3\div 4$) than
the $pn\rightarrow pn\eta$ cross section  which in turn is much larger than
the $pp\rightarrow pp\eta$ cross section (by a factor $6.5$), see Fig. \ref{eta_fig}. 
The above channels for the $\eta$ meson production take the 
$pn\rightarrow pn\eta$ and $pp\rightarrow pp\eta$ reactions 
into account ($N^*(1535)$ is produced with appropriate
cross sections \cite{Teis} in $pp$ and $pn$ collisions), 
but this treatment does not describe properly the reaction 
$pn\rightarrow d\eta$ which is dominant near the $\eta$ threshold. 
At the two lowest incident proton kinetic energies 
$T=1.04$ GeV and $T=1.27$ GeV we add therefore 
the reaction $pn\rightarrow d\eta$ to the $\eta$ production sources 
by a parameterization of the experimental cross section \cite{Calen}. 
At higher incident energies ($T=1.61\div 4.88$ GeV) the 
$pn\rightarrow d\eta$ cross section is not known experimentally 
but it is natural to expect that the enhancement of the cross section 
by the the proton-neutron intial/final state interaction (ISI/FSI) 
in the deuteron becomes negligible 
at high energies. We therefore omit the reaction $pn\rightarrow d\eta$ 
at $T=1.61\div 4.88$ GeV.

The results are presented on Fig. \ref{DLS_fig2}. At incident kinetic 
proton energies of $T=1.04\div 2.09$ GeV dileptons are mainly produced 
from the exclusive reactions mentioned above (exceptions are the 
$\pi^0$ production at  $T=1.85$ GeV and $T=2.09$ GeV and the 
$\eta$ production at $T=2.09$ GeV). At $T=4.88$ GeV this 
procedure strongly underestimates the experimental data. 
The reason is clear: here the inclusive reactions of 
$\pi^0,\eta,\rho,\omega$ production become much larger
than the exclusive ones. As discussed above the resonance model 
provides only exclusive vector meson production cross sections and 
the corresponding dilepton production cross sections.  
In the calculations shown in Figs. \ref{DLS_fig1},\ref{DLS_fig2} we accounted 
for the inclusive cross sections which play a dominant role at high 
incident energies in a simple manner: the ratios of the inclusive/exclusive
cross sections for $\pi^0,\eta,\rho,\omega$ meson production from 
the theoretical predictions of Ref.\cite{ernst} are derived and 
our exclusive cross sections are scaled by the corresponding factors. 
The shape of the experimental curve at $T=4.88$ is then well reproduced. 
At $T=1.04$ GeV and $T=1.27$ GeV we strongly underestimate the 
experimental $pd$ data. This is particularly astonishing 
since the corresponding $pp$ data are reasonably well reproduced. 
The comparison with other available theoretical calculations
\cite{ernst,BCM} shows the following: our $\Delta (1232)$ contribution
is $2\div 2.5$ times smaller than that of Refs. \cite{ernst,BCM}. 
Comparing the $\eta$ contributions to the $pd$ dilepton spectrum 
from \cite{ernst}:\cite{BCM}:[present] yields the following ratios:  
$40:200:6$ at $T=1.04$ GeV and $4:15:8$ at $T=1.27$ GeV. 
However, the large difference of the $\eta$ meson 
contributions at $T=1.04$ GeV does not significantly influence 
the total dilepton yield since the absolute $\eta$ contribution 
is extremely small here. The large difference in the various treatments 
can be attributed to the high momentum tails of the Fermi motion in the 
deuteron which are experimentally not determined, 
and to different $pp:pn$ ratios. The same is probably true at 
$T=1.27$ GeV where the differences concerning the 
$\eta$ contributions ($4:15:8$) are smaller. However, 
this reflects the amount of uncertainty inherent in the theoretical 
description of the $\eta$ production in the $pd$ system around threshold. 
Nevertheless, the isotopic relations and the treatment of the Fermi motion 
can be checked calculating the ratio 
$\sigma(pd\rightarrow \eta X)/\sigma(pp\rightarrow \eta X)$
at two energies $T=1.3~~GeV$ and $T=1.5~~GeV$
where experimental data on these ratios are available \cite{musso}.
Our results are
$$\frac{\sigma(pd\rightarrow \eta X)}{\sigma(pp\rightarrow \eta X)}=(1+5)\cdot 2.0,~~
T=1.3~{\rm GeV}
,~~~\frac{\sigma(pd\rightarrow \eta X)}{\sigma(pp\rightarrow \eta X)}=(1+5)\cdot 0.9,~
~T=1.5 ~{\rm GeV}~,$$
where the first factor originates from the $pn$ isospin relation and 
the second factor is due to the Fermi motion inside the deuteron. 
The corresponding experimental values \cite{musso} of $\approx 10$ and $\approx 5$, 
respectively, demonstrate that the present treatment of the $\eta$ 
production is reasonable.

A second deviation between the present approach and the 
former calculations of Refs.\cite{ernst,BCM} is harder 
to understand. It concerns the contribution of the $\Delta (1232)$ at low 
energies. In the present treatment the dilepton yield from 
the $\Delta (1232)$ in $pd$ reactions is by about a factor 
$2\div 2.5$ smaller than in \cite{ernst,BCM}. Concerning the 
$\Delta (1232)$ there exists no sizable influence of the 
Fermi motion in the deuteron since the reaction is well above 
the kinematical threshold. A comparison of the $pd:pp$ ratios for 
the $\Delta (1232)$ yields approximately 
$(pd:pp)_\Delta \approx 5:1$ in refs.\cite{ernst,BCM} whereas 
we obtain $(pd:pp)_\Delta \approx 3:1$. This latter result is probably 
closer to the required isotopic relation. The simplest way 
to obtain this isotopic relation is the following: 
the deuteron has total isospin $I=0$, the incoming proton has $I=1/2$. 
Therefore, the final $NN\Delta$ system should have total isospin $I=1/2$. 
The isotopic wave function of such a system is unique, i.e. it 
corresponds to $\Delta^{++}$,
 $\Delta^{+}$ and  $\Delta^{0}$ isobars in the proportion
$\Delta^{++}:\Delta^{+}:\Delta^{0} = 3x:2x:1x$. 
Here $x$ is a factor which accounts effectively for the Fermi 
motion of the deuteron constituents. It is only written for the
comparison to the $pp \rightarrow N\Delta$ reaction. 
Let us compare this result to the $\Delta$ contribution 
in $pp \rightarrow N\Delta$ reaction. We have now $\Delta^{++}:\Delta^{+} = 3:1$. 
Radiative decays occur only for $\Delta^{+}$ and $\Delta^{0}$ 
and the radiative widths are equal.
Thus one gets $(pd:pp)_\Delta = 3x:1 \approx 3:1 $ due to $x\approx 1$. 
At $T =1.04;~1.27$ GeV this is an upper limit, i.e. $x <  1$, 
since $NN \rightarrow N\Delta$ is almost on top of the cross section.

In summary the present model reproduces the dilepton production 
in $pd$ collisions at $T=1.61\div 4.88$ GeV rather reasonable. 
At the two lowest energies $T =1.04;~1.27$ GeV  we underestimate
the $pd$ data (probably due to an underestimation of the $\eta$ 
contribution). At these energies an underestimation which is, however, 
less pronounced, was also observed in \cite{ernst}. It should 
be noted that for the $pp$ reactions the present results and those of 
 \cite{ernst,BCM} coincide more or less. In all cases the 
theoretical calculations reproduce the corresponding DLS data 
reasonably well. Hence the dilepton production on the deuteron 
turns out to be rather involved at subthreshold energies due to 
strong ISI/FSI effects. The $pd$  
system is therefore only of limited use to check isospin relations 
of the applied models. Another important result is the fact that 
the scenario of large off-shell $\omega$ contributions from the 
$N^*(1535)-N\omega$ decay is consistent with the 
available $pp$ and $pd$  dilepton data.

\section{Decoherence as a medium effect}
In this Sect. we discuss an in-medium modification of the cross
section ${NN\rightarrow e^{+}e^{-}X}$ which is connected with the decoherence of
vector mesons propagating in a hot and dense nuclear medium. 
In refs. \cite{resdec,krivo02}, radially excited $\rho $- and 
$\omega $-mesons were introduced in the transition form factors $RN\gamma $ to
ensure the correct asymptotic behavior of the amplitudes in line with the
quark counting rules. Thereby we required a destructive interference
between the members of the vector meson families away from the poles of 
the propagators, i.e. the meson masses. In a dense medium the environment 
of the vector mesons can be regarded as a heat bath. Usually the 
different scattering channels of the interaction with a heat 
bath, i.e. the surrounding nucleons and pions, are summed up 
decoherently since the various channels acquire 
large uncorrelated relative phases. 
In such a case, the coherent 
contributions to the probability are random and cancel each other. 
We have in a sense macroscopically different intermediate states which do 
not interfere since small perturbations result in 
macroscopically large variations of the relative phases. 
The interaction of the vector mesons with the 
surrounding particles should therefore break up the 
coherence between the corresponding 
amplitudes for the dilepton production. The break up of the 
destructive interference results in an increase of the
total cross sections at low dilepton masses. In the following 
we want to investigate if the decoherence effect can explain 
the enhancement observed in the dilepton spectra at the BEVALAC 
experiment (DLS puzzle). Below we put this idea on a more 
quantitative basis.

\subsection{In-medium modification of the transition form factors}
The decay widths of nucleon resonances with spin $J=l+\frac{1}{2}$ and 
mass $\mu$ into a nucleon with mass $m_N$ and a dilepton pair with mass $M$ are 
described by Eqs.(3)-(6). These widths are proportional to squares of the
magnetic ($M$), electric ($E$), and Coulomb ($C$) form factors $G_{T}^{(\pm
)}(M^{2})$ ($T=M,E,C$). In the eVMD model, the transition 
form factors $RN\gamma $ are written as 
\begin{equation}
G_{T}^{(\pm )}(M^{2})=\sum_{k}{\cal M}_{Tk}^{(\pm )}.  \label{E2}
\end{equation}
The sum runs 
over the ground state and excited $\rho $- and $\omega $-mesons. The amplitude 
\begin{equation}
{\cal M}_{Tk}^{(\pm )}=h_{Tk}^{(\pm )}\frac{m_{k}^{2}}{m_{k}^{2} - i m_{k}\Gamma_{k}-M^{2}}
\end{equation}
describes the contribution from the $k$-th vector meson to the 
type-$T$ decay width. The quark counting rules 
\cite{VZ,Matveev:1973ra} predict the following
asymptotics for the covariant form factors of $J\geq \frac{3}{2}$ nucleon
resonances: 
\begin{eqnarray}
-lG_{E/M}^{(\pm )}(M^{2})\simeq G_{M/E}^{(\pm )}(M^{2})
\sim O(\frac{1}{(-M^{2})^{l+1}})
\quad,\quad G_{C}^{(\pm )}(M^{2})\sim O(\frac{1}{(-M^{2})^{l+2}})~.
\end{eqnarray}
These relations provide constraints to the residues $h_{Tk}^{(\pm )}$ 
and imply a destructive interference
between the different members of the vector meson families. 

For spin $J=\frac{1}{2}$ resonances one obtains the following asymptotics 
\begin{eqnarray}
G_{E/M}^{(\pm )}(M^{2})\sim O(\frac{1}{(-M^{2})^{2}})
\quad,\quad 
G_{C}^{(\pm )}(M^{2})\sim O(\frac{1}{(-M^{2})^{3}})~.
\end{eqnarray}

In the case of a full decoherence the vector meson contributions to the cross
section ${NN\rightarrow e^{+}e^{-}X}$ which run over nucleon resonances
must be summed up decoherently. This leads to the replacement 
\begin{equation}
|\sum_{k}{\cal M}_{Tk}^{(\pm )}|^{2}\rightarrow \sum_{k}|{\cal M}%
_{Tk}^{(\pm )}|^{2}~.  \label{E7}
\end{equation}
As a consequence, total decoherence will result in an 
enhancement of the resonance contributions due to the 
presence of the medium. The prescription (\ref{E7}) refers to 
the limit of full decoherence, i.e. collisions 
with nearest neighbors occur always before the dilepton emission. 
However, in reality both, the density and the meson wavelengths are 
finite and thus it is necessary to 
have a relation for the decoherence effect which is valid in an intermediate 
regime for densities and the meson wavelengths. 
The basic assumption is that each of the propagated vector mesons
radiates $e^{+}e^{-}$ pairs coherently up to its first collision with a
nucleon (or generally a hadron) and incoherently afterwards. This leads to the
destruction of the coherence of one meson with the others which, by themselves,
may still form a coherent state. The problem receives at this stage 
a combinatorial character.

The decay probability for a resonance at 
distance $l_{C}$ in the interval $dl_{C}$ equals 
\begin{equation}
dW_{D}(l_{D})=e^{-l_{D}/L_{D}}\frac{dl_{D}}{L_{D}}~.
\end{equation}
The decay length for a resonance with lifetime $T_{D}$
equals $L_{D}=v\gamma T_{D}$, where $T_{D}=1/\Gamma $, $\Gamma $ being the
total vector meson vacuum width.
The collision probability at a distance $l_{C}$ in the interval $%
dl_{C}$ equals 
\begin{equation}
dW_{C}(l_{C})=e^{-l_{C}/L_{C}}\frac{dl_{C}}{L_{C}}~.
\end{equation}
The collision length $L_{C}$ is defined by the expression 
\begin{equation}
L_{C}=\frac{1}{\rho_B\sigma }  \label{coll}
\end{equation}
where $\sigma $ is the total $VN$ cross section and $\rho_B$ is the
nuclear density. 
\begin{figure}[h]
\begin{center}
\leavevmode
\epsfxsize = 10cm
\epsffile[50 50 510 470]{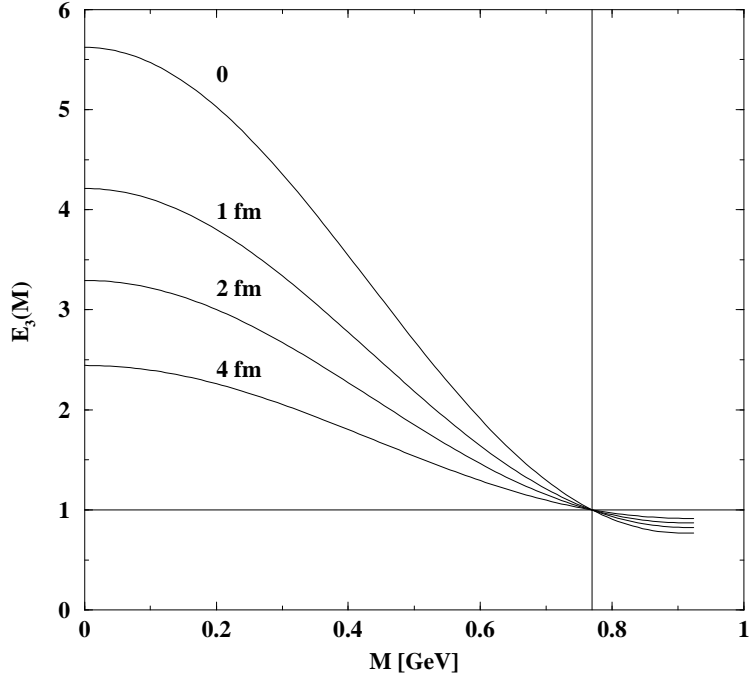}
\end{center}
\caption{The enhancement factor $E_{C}(M)$ for the spin-1/2 
$\Delta \rightarrow Ne^+e^-$ 
Coulomb transition due to the decoherence between the 
$\rho$-mesons in the medium, 
estimated within the eVMD model for different values of the mean free path 
$L_C$ of the $\rho$-mesons in the medium. Three $\rho$-mesons interfere.}
\label{en3_fig}
\end{figure}
The meson decay takes place before the first collision provided that 
$0<l_{D}<l_{C}$, so the probability of the coherent decay equals
\begin{eqnarray}
w =\int_{0}^{+\infty }\frac{dl_{C}}{L_{C}}e^{-l_{C}/L_{C}}%
\int_{0}^{l_{C}}\frac{dl_{D}}{L_{D}}e^{-l_{D}/L_{D}}  = \frac{L_{C}}{L_{C}+L_{D}}~.
\label{decprob1}
\end{eqnarray}
All mesons have in general different values
$L_{D}$ and $L_{C}$ and thus the coherent decay probabilities are
different as well. Therefore below the index $k$ is attached 
to the decay and scattering lengths 
and to the coherent decay probabilities. In order to account for the decoherence 
one should make the replacements 
\begin{equation}
\left| G_{T}^{(\pm)}(M^{2})\right|^{2} \rightarrow E_{T}^{(\pm)}(M^{2},{\bf Q}^{2})
\left| G_{T}^{(\pm )}(M^{2})\right|^{2}
\end{equation}
in Eqs.(5) and (6). The enhancement factor $E_{T}(M^{2},{\bf Q}^{2})$ is
given by 
\begin{eqnarray}
E_{T}^{(\pm)}(M^{2},{\bf Q}^{2}) = \left( \prod_{k}w_{k}|\sum_{k}{\cal M}_{Tk}^{(\pm)}|^{2}
+ \sum_{l}(1-w_{l})\prod_{k\neq l}w_{k}(|{\cal M}_{Tl}^{(\pm )}|^{2}
+|\sum_{k\neq l}{\cal M}_{Tk}^{(\pm )}|^{2})\right.
\nonumber \\
\left. +...+\prod_{l}(1-w_{l})\sum_{k}\left| {\cal M}_{Tk}^{(\pm )}\right|
^{2}\right) /\left( |\sum_{k}{\cal M}_{Tk}^{(\pm )}|^{2}\right). 
\label{coh3}
\end{eqnarray}
It depends on square of the space like part ${\bf Q}$ of the vector meson
momentum through Eq.(\ref{coll}). The first term in Eq.(\ref{coh3}) in the
numerator corresponds to the probability that all $\rho $-mesons radiate the
dilepton pairs coherently. The second term corresponds to the probability
that the $l$-th meson decays to the dilepton pair after its first collision,
while the other mesons radiate before the first collision. Finally, the last
term corresponds to the probability for an incoherent radiation of all vector
mesons. Each term in Eq.(\ref{coh3}) contains the squares of the
amplitudes ${\cal M}_{Tk}^{(\pm )}$ according to the proper interference
pattern. If the probability for the coherent radiation equals $w_{k}=1$, 
i.e. the collision length $L_{C}$ is infinite like in the vacuum, then 
the vacuum result is recovered $E_{T}^{(\pm )}(M^{2},{\bf Q}^{2})=1.$ 
If the collision length goes to zero $w_{k}=0$ (full decoherence) 
prescription (\ref{E7}) is valid. 
In the case of isospin $I=\frac{1}{2}$ resonance decays,  
Eq.(\ref{coll}) takes also the decoherence between $\rho$- and $\omega$-mesons 
into account. 

In order to illustrate the effect of the enhancement factor, we consider the
Coulomb form factor for a spin-1/2 $\Delta $-resonance where the
formulae are simplest. According to the minimal eVMD 
three $\rho $-mesons are needed to ensure the correct
asymptotics of the transition form factors, 
i.e. the ground-state and the excited $\rho $(1250) and $\rho $(1450). 
Let us take $L_{D}\approx T_{D} = 1/\Gamma,$ $w_{1}=w_{2}=w_{3}$
and vary the collision length $L_{C}$ from $0$ (total decoherence) to 
$\infty $ (total coherence). The decoherence factor is plotted on Fig. 
\ref{en3_fig} as a function of the running mass $M$ in the no-width 
approximation for the $\rho$-mesons. As can be seen from Fig. \ref{en3_fig}, 
the decoherence will generally lead to an enhancement of the dilepton yield 
in the low-mass region below the $\rho $-peak. 

It should be noted that a similar effect exists for the dilepton decays of the mesons. 
Such decays have also constraints from the quark counting rules on the 
asymptotic behavior of the transition form factors. 
The decay modes $P\rightarrow e^{+}e^{-}\gamma \;$where$\;P=\pi ,\;\eta \;$
and $\rho ^{0}\rightarrow e^{+}e^{-}\pi ^{+}\pi ^{-}$ have monopole form
factors in the amplitudes. To obtain a monopole form factor it is sufficient to
consider only a single $\rho $-meson. In this case no enhancement occurs, i.e. 
$E(M^2,{\bf Q}^2) \equiv 1$. The decay modes $V\rightarrow e^{+}e^{-}P,$ $\eta
\rightarrow e^{+}e^{-}\pi ^{+}\pi ^{-},$ $\rho ^{0}(\omega )\rightarrow
e^{+}e^{-}\pi ^{0}\pi ^{0},\ $with dipole form factors require the existence
of at least two $\rho $-states. In such a minimal case, these modes are
enhanced. However, the decays of the last type are non-dominant 
and their enhancement is not taken into account in the simulations.

\subsection{Restoration of coherence in the soft-dilepton limit}
Physically, if many nucleons appear on the scale of the mesonic 
wavelength, the scattering process must have a coherent character 
with respect to clusters formed by the surrounding nucleons. 
In such a case Eq. (\ref{coll}) does not apply any more. 
The eVMD model can also be used for the  description of the 
diagonal electromagnetic form
factors. When $M={\bf Q}^{2}=0$, the diagonal form factors, e.g. of the 
nucleon, measure the total electric charge (through $G_{E}$). 
The nucleon charge must be counted in the same way as in the vacuum 
which leads to the requirement 
$E_{E}(M^{2},{\bf Q}^{2})=1$ at $M={\bf Q}^{2}=0$ for the enhancement factor
of the nucleon electric Sachs form factor. Since the in-medium 
behavior of vector mesons does not depend on their origin 
(emission from nucleons or nucleon resonances), the constraints to
the diagonal and the transition form factors must be identical. Hence, in the
soft-dilepton limit the coherence must be restored.

Eq. (\ref{coll}) is the leading term when the 
density approaches zero. The condition for a fully 
decoherent scattering of particles propagating through 
a medium is the dilute gas limit. It means that sequential 
scattering processes are statistically 
independent. In terms of a scattering length, the dilute gas 
limit corresponds to the requirement that no additional scattering 
centers appear inside the wave zone of the scattered particle. 
In the present case this area can be estimated by a 
sphere of radius $r$ which is of the order of  the
meson wave length, $r\sim \lambda$. 
In the low density limit this condition is satisfied and 
Eq. (\ref{coll}) is applicable.

In the following we intend to derive a modified 
expression for the collision length which
describes {\it qualitatively} also the intermediate and high density regime 
and provides the restoration of coherence in the soft-dilepton limit. 
The scattering has a coherent character 
if many scattering centers appear on the scale of the particle's 
wavelength $\lambda $. For a coherent scattering process on a cluster 
which consists of $Z$ individual scattering centers the cross section 
is given by 
\begin{equation}
\sigma _{Z}\sim Z^{2}\sigma~.
\end{equation}
where $\sigma $ is the cross section for a single 
scattering center ($Z=1$). If one assumes - as usually done - 
that the scattering centers are homogeneously distributed according 
to the density $\rho_B$,  the probability to find a cluster with $Z$ 
scattering centers inside of a volume $V$ is given by the Poisson distribution 
\begin{equation}
P_{Z}=\frac{\alpha ^{Z}}{Z!}e^{-\alpha }~.
\end{equation}
Here $\alpha =\rho_B V$ is the average number of scattering centers in the volume 
$V$. Coherent scattering takes place on clusters inside of a sphere of
radius $r\sim \lambda$. The average cross section for the scattering on 
clusters equals 
\begin{equation}
\sigma _{\rm clus}\sim \sigma \sum_{Z=0}^{+\infty }Z^{2}\frac{\alpha ^{Z}}{Z!}%
e^{-\alpha }=\sigma \alpha (1+\alpha )~.  \label{10}
\end{equation}
The average number of scattering centers inside a single cluster is 
\begin{equation}
\bar{Z}=\sum_{Z=0}^{+\infty }Z\frac{\alpha ^{Z}}{Z!}e^{-\alpha }=\alpha ~.
\label{11}
\end{equation}
The ratio between Eqs. (\ref{10}) and (\ref{11}) provides now the 
effective cross section for the scattering on a single scattering center: 
\begin{equation}
\sigma _{\rm eff}\sim \sigma (1+\alpha )~.  
\label{20}
\end{equation}
In the case of decoherent scattering the above arguments lead to the 
relations $\sigma _{Z}\sim Z\sigma ,$ $\sigma _{\rm clus}\sim \sigma \alpha$, 
and $\sigma_{\rm eff}\sim \sigma$.

In relativistic heavy ion reactions the masses and momenta 
which occur in hadronic scattering processes are usually large 
and thus quantum interference effects do not play a significant role. 
But here we are interested in the soft limit of the vector meson 
propagation and thus one has to account for quantum effects. From 
scattering theory one knows that radiation takes place if the 
asymptotic regime $\sim 1/r$ starts for the wave function of the 
scattered particle outside the wave zone. When scattered on a cluster, 
the incident particle can hit new scattering centers inside the wave zone 
and in this case radiation is assumed not to be formed. This 
means that a discrete scattering process can only take place on those clusters 
which leave the wave zone of the scattered particle unblocked, 
i.e. free from new scattering centers. The probability to find such a 
configuration can be estimated by the Poisson law: 
\begin{equation}
P_{\rm unblocked}\sim e^{-\alpha }~.
\end{equation}
The value of $P_{\rm unblocked}$ is the probability that no 
additional scattering centers exist inside the wave zone which 
we consider to be simply a region around the scattering cluster of 
the same volume $V$. The collision probability is then proportional to the effective 
cross section $\sigma _{\rm eff}$ multiplied by the probability $P_{\rm unblocked}$
for an unblocked wave zone. The modification of Eq. (\ref{coll}) 
is now straightforward: 
\begin{equation}
L_{C}\sim \frac{e^{\alpha }}{\rho_B \sigma (1+\alpha )}~,  \label{collH}
\end{equation}
with $\alpha =\rho_B \frac{4\pi }{3}\lambda ^{3}$. 
Expression (\ref{collH}) has finally the desired features. 
In the low density limit one obtains $\alpha \rightarrow 0$ and thus 
expression (\ref{coll}) is recovered. In the long wave 
limit $\alpha \rightarrow \infty$, $L_C \rightarrow \infty$, $w \rightarrow 1$, and so 
the full coherence is restored. Note that the function $%
e^{\alpha }/(1+\alpha )$ is a monotonously increasing function. 

The wavelength $\lambda $ is inverse proportional to the center-of-mass
momentum of the vector meson and the cluster, 
\begin{equation}
\frac{1}{\lambda }\sim p^{*}(\sqrt{s},M,\bar{m})
\end{equation}
where $\bar{m}^{2}=(\sum_{i=1}^{Z}p_{i})^{2}$, $p_{i}$ are the four momenta of 
the nucleons in the cluster. Here $s=(P+\sum_{i=1}^{Z}p_{i})^{2}$ and $P$ is the vector 
meson momentum, $P^{2}=M^{2}$. In the local rest frame of the cluster, i.e. 
the center-of-mass frame of its constituents, the vector meson momentum 
is given by
\begin{equation}
p^{*}(\sqrt{s},M,\bar{m})=\frac{\bar{m}}{\sqrt{s}}|{\bf Q}_{\rm clus}|
\end{equation}
where $s=M^{2}+2P_{0}\bar{m}+\bar{m}^{2}$ and 
$P_{0}=\sqrt{M^{2}+{\bf Q}_{\rm clus}^{2}}.$ In order to obtain an infinite wavelength 
$\lambda =\infty$ one has to require that the vector meson momentum vanishes 
simultaneously in the rest frames of all clusters, ${\bf Q}_{\rm clus}^{2}=0$. 
This is, however, only possible if the condition $M={\bf Q}^{2}=0$ is 
fulfilled. Thus,  at finite density a 
full restoration of the coherence can only take place for $M={\bf Q}^{2}=0$. 
This condition appears quite reasonable, since a vector meson at rest 
with $M \neq 0$ and ${\bf Q}^2=0$ can still collide with the surrounding nucleons
due to the Fermi motion and/or motion caused by a finite temperature.

It is interesting to note that $L_{C}\rightarrow \infty $ both, 
at $\rho_B\rightarrow 0$ and $%
\rho_B \rightarrow \infty $. This implies the full restoration of coherence at
finite $\lambda $ for both, small and infinite densities. 
For large clusters ($Z \rightarrow \infty$) $\bar{m}\sim Zm$ becomes dominant over 
$M$ and $P_{0}$, and so $p^{*}(\sqrt{s},M,\bar{m})\rightarrow |{\bf Q}_{\rm clus}|.$ 
The c.m. velocity ${\bf v}_{\rm clus}$ of a large cluster relative 
to the matter rest frame vanishes as ${\bf v}_{\rm clus}^{2}\sim 1/Z$. 
It follows that $|{\bf Q}_{\rm clus}| \rightarrow |{\bf Q}|$ 
and $1/\lambda  \rightarrow |{\bf Q}|.$
For a single scattering center, $\bar{m} = m$, and the
wavelength $\lambda $ is determined by the momentum $p^{*}(\sqrt{s},M,m)$
averaged over the nucleon velocity distribution in the matter.

In deriving Eq.(\ref{collH}), we neglected the dependence of $\lambda$ on $Z$.
Although very qualitative, Eq.(\ref{collH}) provides the desired behavior of the
decoherence factors in the soft-dilepton limit. It leads to $L_{C}\rightarrow
\infty ,$ $w_{k}\rightarrow 1,$ $E_{T}^{(\pm )}(M^{2},{\bf Q}%
^{2})\rightarrow 1$ at $\lambda \rightarrow \infty $ $(M,{\bf Q}%
^{2}\rightarrow 0)$, so that vector mesons with $M,{\bf Q}^{2}\rightarrow 0%
{\bf \ }$propagate in a dense medium coherently. The decoherence becomes
generally weaker with increasing $\lambda $. 

The requirement of a restoration of coherence in the soft-dilepton limit
 follows directly from charge conservation. It is of principle importance 
but has no immediate practical implications for the 
description of experimental spectra. 
The experimental filters cut the dilepton spectra at low values of $M$ and 
thus this limit is presently not accessible. We do not discuss here possible
effects of the mass dependence of the decay time through equation
$T_{D} = 1/\Gamma(M^2)$ or through Eq.(\ref{resdec}). Note also that the
meaning of the cross section entering the collision length $L_C$ becomes unclear
when $M$ falls below the two-pion threshold (for $\rho$-mesons), so the above 
discussion is restricted to the case of massless pions.

\section{The QMD transport model}
Heavy ion reactions are described within the framework of the
Quantum Molecular Dynamics (QMD) transport model \cite{ai91}. We
extended our QMD transport code \cite{uma97} in order to include all
nuclear resonances with masses below 2 GeV. These are altogether 
11 $N^*$ and 10 $\Delta$ resonances. The corresponding masses 
and decay widths are listed in Tables \ref{nstar_tab} and \ref{delta_tab}. 
For the description of the dilepton production through baryonic 
resonances, respectively the $\rho$ and $\omega$ production in $NN$ and 
$\pi N$ reactions, only the well established ($4*$) resonances listed 
by the PDG \cite{pdg} are taken into account. This corresponds to the 
same set of resonances which was used in \cite{resdec,omega02} for 
the description of vector meson and dilepton production. $\Gamma_{\rm tot}$, 
the $N\rho$ and $N\omega$ widthes given in brackets as well as the decay 
widthes of the other decay channels are taken from \cite{Bass} and used for the reaction 
dynamics. 
\begin{figure}[h]
\begin{center}
\leavevmode
\epsfxsize = 12cm
\epsffile[20 50 550 390]{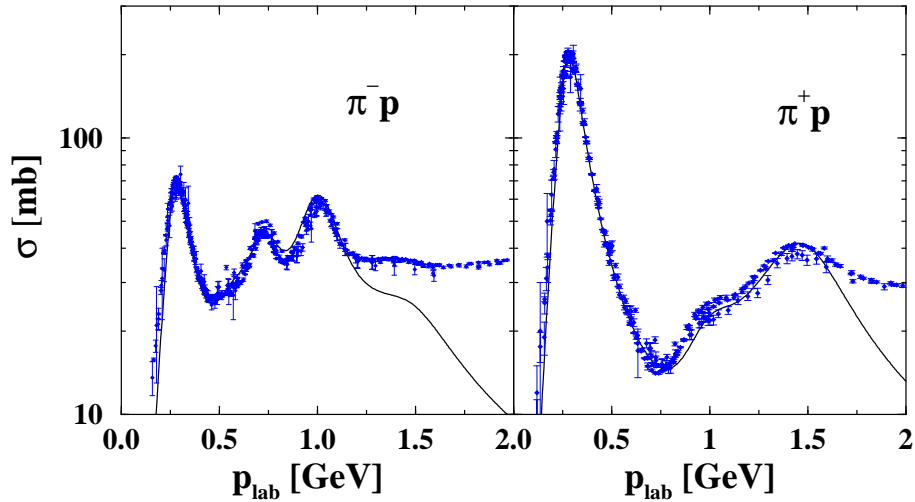}
\end{center}
\caption{Inclusive $\pi^- p$ and $\pi^+ p$ cross sections obtained 
by the sum over all resonances which are taken into account in the present 
description (see Tables \protect\ref{nstar_tab} and \protect\ref{delta_tab}). 
Data are taken from \protect\cite{pdg}.
}
\label{pion_fig}
\end{figure}
As in the previous calculations \cite{uma97} we take the iso-spin dependent
production cross sections  $\sigma^{NN\rightarrow NR}$ for the
$\D (1232)$ and the $N^* (1440)$ resonances from \cite{Hu94}. These
cross sections were determined within the framework of a
one-boson-exchange model. For the higher lying resonances 
parameterizations for the production cross-section are taken from  
different sources \cite{Bass,Teis}. The following types of 
baryon-baryon collisions are included:
all elastic channels, reactions of the type $NN\rightarrow NN^*$,
$NN\rightarrow N\D^*$, $NN\rightarrow\D_{1232}N^*$,
$NN\rightarrow\D_{1232}\D^*$ and $NR\rightarrow NR^\prime$, where $\D^*$
denotes all higher lying $\D$-resonances. 
Elastic scattering is considered on the same footing for all the particles
involved. Matrix elements for elastic reactions are assumed to 
be the same for nucleons and nucleonic resonances. Thus elastic $NR$ and 
$RR$ cross sections are determined from the elastic $pp$  or
$np$ cross sections, depending on the total charge. 
Inelastic collisions are considered according to the expression \cite{Bass}
\begin{equation}
\sigma_{1,2\rightarrow 3,4} \sim \frac{\left<p_f \right>} {p_i s} \left|{\cal M}
(m_3, m_4)\right|^2
\label{cr1}
\end{equation}
$p_i$ and $<p_f>$ are
the momenta of incoming and outgoing particles in the center of mass
frame. In the case that final states are resonances, 
the phase space has to be averaged over the corresponding spectral function 
\begin{equation}
<p_f> =
\int p(\sqrt{s},m_N,\mu)~ dW_{R^\prime}(\mu) 
\label{cr3}
\end{equation}
with $dW_{R^\prime}$ given by the corresponding 
Breit-Wigner distribution (\ref{BW}). 
In the general case that both final states in eq. (\ref{cr1}) are 
resonances the averaging of  $p_f$ is performed over both  resonances 
\begin{equation} 
<p_f> = \int p(\sqrt{s}, \mu, \mu^\prime) 
~dW_{R}(\mu) ~dW_{R^\prime}(\mu^\prime )
\label{momenta}
\end{equation}
The integrations are performed over kinematically defined limits. 
$\cal M$ in Eq. (\ref{cr1}) is the matrix element of the cross-section and 
the proportionality sign accounts for
possible overall (iso-)spin coefficients. For most of the cases we
use expressions for the matrix elements from Ref. \cite{Bass}. However,
parameterizations of the  matrix elements are given in Ref. \cite{Teis}, 
we make use of these expressions. This is in particular the case for reactions 
where resonances contribute to the dilepton 
yield (see Tables \ref{nstar_tab} and  \ref{delta_tab}). 
E.g. the cross-section for the reactions $NR\rightarrow NR^{\prime}$ is
determined from the known channels $NN\rightarrow NR$ and
$NN\rightarrow NR^{\prime}$ by 
\begin{equation}
\sigma_{NR\rightarrow NR^\prime} = I\frac{0.5
(|{\cal M}_{NN\rightarrow NR}|^2 + |{\cal M}_{NN\rightarrow NR^\prime}|^2)
2(2J_{R^\prime} + 1)}{16\pi p_i s}<p_f> ~.
\label{cr2}
\end{equation}
In eq. (\ref{cr2}) $I$ is an isospin coefficient, 
depending on the resonances' types, and  $J_{R^\prime}$ 
denotes the spin of $R^\prime$. 

For all resonances 
we use mass-dependent widths in expressions (\ref{cr2}-\ref{momenta}), 
namely
\begin{equation} 
\Gamma (\mu ) = \Gamma_R\left(\frac{p}{p_r}\right)^3
\left(\frac{p_r^2 + \delta^2}{p^2 + \delta^2}\right)^2~~.
\label{gam1}
\end{equation}
In (\ref{gam1}) $p$ and $p_r$ are the c.m. momenta of the pion 
in the resonance rest frame evaluated at the running and the 
resonance pole mass, respectively. $\delta=0.3$ is chosen for the 
$\D_{1232}$ and $\delta = \sqrt{(m_R - m_N - m_\pi)^2 + \Gamma^2/4}$ for the
rest of the resonances. The inclusive $\pi^- p$ and  $\pi^+ p$ 
cross sections are shown in Fig. \ref{pion_fig}. The fit to the data 
including the sum over all resonances is of similar quality as in refs. 
\cite{Teis,Bass} and reproduces the absorption cross section up to 
pion laboratory momenta of 1-1.5 GeV. At higher energies string excitations 
start to play a role \cite{Bass}. 

Backward reactions, e.g. $NR\rightarrow NN$, are treated by detailed balance
\begin{equation} 
\sigma_{3,4\rightarrow 1,2} \sim \frac{|p_{1,2}|^2}{|p_{3,4}|^2}
\sigma_{1,2\rightarrow 3,4}
\end{equation}
where the proportionality sign is due to overall (iso-)spin factors.
The expressions for the momenta of incoming (outgoing) particles are 
calculated according to (\ref{cr2},\ref{momenta}), respectively. 

Pion-baryon collisions are standardly treated as two-stage
processes, i.e. first the pion is absorbed by a nucleon or a 
baryonic resonance forming a new resonance state with subsequent 
decay. The pion absorption  by nucleons is treated 
in the standard way \cite{uma97,Teis,Bass} 
and the pion absorption by resonances is proportional to the 
partial decay width of the reverse process \cite{Teis}
\begin{equation} 
\sigma_{\pi R\rightarrow R^\prime} = \frac{2J_{R^\prime}+1}
{(2S_a + 1)(2S_b + 1)}\frac{4\pi}{p_i^2}
\frac{s(\Gamma_{R^\prime\rightarrow R\pi})^2}
{(s-m_{R^\prime}^2)^2 + s\Gamma_{R^\prime}^2}~~.
\label{piabs1}
\end{equation}
The decay of baryonic resonances is treated as proposed in 
\cite{pratt96,aichelin97,mosel01}, i.e. the resonance life 
time is given by the spectral function 
\beq
\tau_R (\mu)  = 4\pi \mu \frac{dW_{R}(\mu)}{d\mu^2}
\label{resdec}
\eeq
Here we use constant widths when considering resonance decays. 
The decay channels which are 
taken into account are listed in Tables \ref{nstar_tab}, \ref{delta_tab} 
together with their corresponding branching ratios. For the mass systems under 
consideration pion multiplicities are reasonably well reproduced
 by the present description. E.g. inclusive $\pi^+$ cross sections in 
$C+C$ reactions were recently measured by the KaoS Collaboration 
\cite{sturm01} and the experimental results can be reproduced by 
the present description within error bars.

Concerning the $\eta$ the fit of \cite{sibirtsev97} is in good agreement 
with the exclusive $pp\rightarrow pp\eta$ 
production data from COSY \cite{cosy11} 
around threshold. Thus in this case we apply the cross section from 
 \cite{sibirtsev97} and neglect the  $\eta$ production through 
resonances. As a consistency check we compared the direct $\eta$ 
production by the process $NN\rightarrow NN \eta $ to that of  
$NN\rightarrow RN\rightarrow NN \eta$ and found that the two 
production mechanisms lead to almost identical $\eta$ yields in 
heavy ion reactions. However, to avoid double counting only one 
of the channels should be included. In line with experimental 
data \cite{faeldt} for the $\eta$ an iso-spin factor of 
\begin{equation}
\sigma(pn\rightarrow pn\eta)$  = 6.5 $\sigma(pp\rightarrow pp\eta)
\end{equation}
is assumed. 
\section{Dilepton production in heavy ion reactions}
\subsection{Standard treatment}
With this input QMD transport calculations for $C+C$ 
and $Ca+Ca$ reactions at 1.04 AGeV are performed. 
First we discuss the results obtained without any additional medium 
effects concerning the dilepton production. For the nuclear mean field 
a soft momentum dependent Skyrme force (K=200 MeV) is used \cite{ai91}
which provides also a good description of the subthreshold $K^+$ 
production in the considered energy range \cite{fuchs01}. 
The reactions are treated as minimal bias collisions with maximal impact 
parameters $b_{\rm max}=5(8)~{\rm fm}$ for $C+C (Ca+Ca)$. 

\begin{figure}[h]
\begin{center}
\leavevmode
\epsfxsize = 14cm
\epsffile[20 50 550 390]{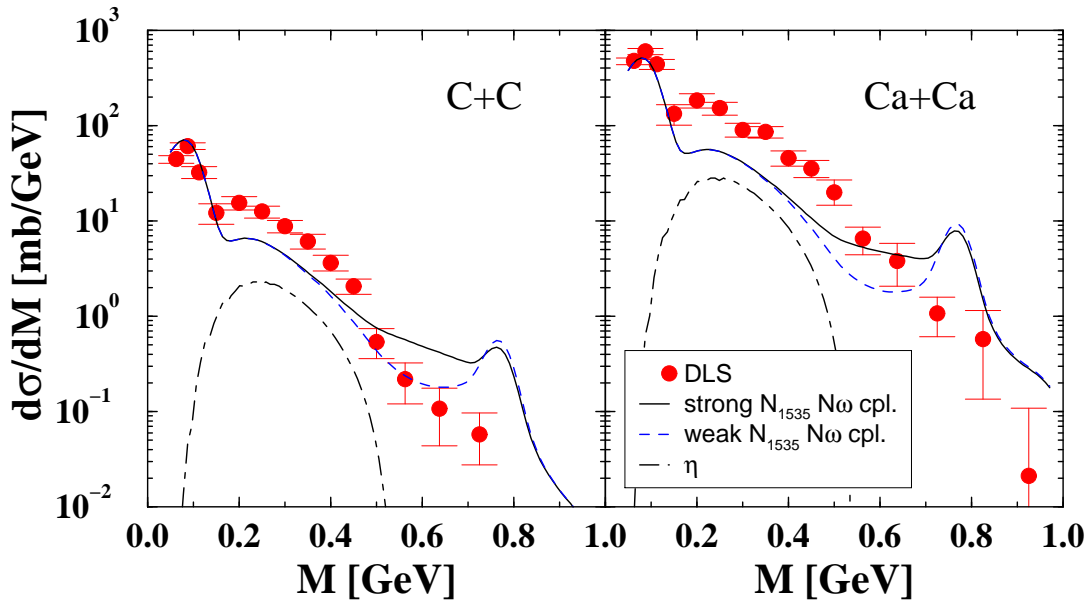}
\end{center}
\caption{The dilepton spectrum in $C+C$ and $Ca+Ca$ reactions is 
compared to the DLS data \protect\cite{DLS}. The calculations are 
performed with a strong, respectively a weak $N^*(1535)-N\omega $ 
coupling.
}
\label{DLS_AA_fig1}
\end{figure}

In Fig. \ref{DLS_AA_fig1} the results are compared to the DLS 
data. The acceptance filter functions provided by the 
DLS Collaboration are applied and the results are smeared over the 
experimental resolution of $\Delta M = 35$ MeV.  The calculations are 
performed within the two scenarios discussed in Sec. II, namely a 
strong  $N^*(1535)-N\omega $ coupling as implied by the original fit to 
the available photo-production data \cite{omega02} and a weaker coupling 
which can be enforced by a different choice of input parameters. In the 
first case strong off-shell $\omega$ contributions appear which are also 
visible in the dilepton spectrum at invariant masses below the 
$\omega$ peak. In the mass region between $0.4\div 0.8$ GeV the two 
scenarios yield significantly different results. The rest of the 
spectrum is practically identical except from the height of the $\omega$ 
peak itself. As discussed in connection with the elementary cross 
sections the $\omega$ contribution from the $N^*(1535)$ is suppressed 
at the  $\omega$ pole in the strong coupling scenario and thus the 
total  $\omega$ peak is slightly lower. The comparison of the transport calculations 
with the DLS data is here not completely conclusive: The lighter $C+C$ 
system would favor the weak  $N^*(1535)-N\omega $ coupling scenario 
whereas the $Ca+Ca$ reactions are better described by the strong coupling. 
In the low mass region ($M=0.1\div 0.5$ GeV) we observe an underestimation 
of the DLS spectra  by a factor of $2\div3$. Thus in 
the present approach the underestimation of the DLS data 
is somewhat smaller than observed in the previous works of 
\cite{ernst} and \cite{BCRW98}. One reason for this is a larger 
$\eta$ contribution which is probably due to 
the iso-spin factor of 6.5 for the $np\rightarrow np\eta$ channel 
(compared to a factor of 2.5 used in \cite{BCRW98,BCM}). Other differences 
to the previous treatments \cite{ernst,BCRW98} are the following: 
In ref. \cite{BCRW98} the vector meson production 
was described by parameterizations of the $NN$ and $\pi N$ production channels 
while in the present approach these reactions run solely over the excitation of 
intermediate nuclear resonances. In \cite{ernst,BCRW98} only the 
$\Delta(1232)\rightarrow Ne^+ e^-$ Dalitz decay has explicitely been 
included. In addition, the decays of the nucleon resonances 
into vector mesons were treated till recently 
in the non-relativistic approximation \cite{BCM,pirner} 
and usually only one transition form factor was taken 
into account. From counting the independent helicity amplitudes it is clear that
a phenomenologically complete treatment requires
three transition form factors for spin $J \geq 3/2$ nucleon resonances and 
two transition form factors for spin-1/2 resonances.
Earlier attempts to derive a complete phenomenological 
expression for the dilepton decay of 
the $\Delta(1232)$ were not successful (for a discussion see \cite{krivo01}).
Despite of the details which differ in the various transport 
calculations (we included significantly more decay channels and apply an 
improved description of the baryonic resonance decays) the 
present results confirm qualitatively the underestimation of the 
DLS data at invariant masses below the $\rho/\omega$ peak \cite{ernst,BCRW98}.

A deviation to the results of \cite{ernst} and \cite{BCRW98} appears in the 
vicinity of the $\omega$ peak. Even after averaging over the experimental 
resolution the present results show a clear peak structure around 0.8 GeV 
which is absent in \cite{ernst,BCRW98}. However, in \cite{BCRW98} absorptive 
channels (e.g. $N\omega\rightarrow N\pi$ \cite{cassing99}) have been 
included which lead automatically to a collisional broadening of the 
in-medium vector meson width. Such a collision broadening is not 
included in the results shown in Fig. \ref{DLS_AA_fig1} but will 
separately be discussed in the next subsection. 
With respect to the  UrQMD calculations of \cite{ernst} our 
approach is in principle similar since vector mesons are produced through 
the excitation of nuclear resonances. However, in \cite{ernst}  the naive 
VMD was applied to treat the mesonic decays and the treatment is more qualitative, 
i.e. couplings were not particularly adjusted in order to describe 
$\rho$ and $\omega$ cross section as it was done in \cite{krivo02,omega02}. 
E.g. in  \cite{ernst} only the $N^*(1900)\rightarrow N\omega$ 
decay mode was taken into account which leads presumably to a significant 
underestimation of the $NN\rightarrow NN\omega$ cross section. 
\begin{figure}[h]
\begin{center}
\leavevmode
\epsfxsize = 14cm
\epsffile[20 50 550 390]{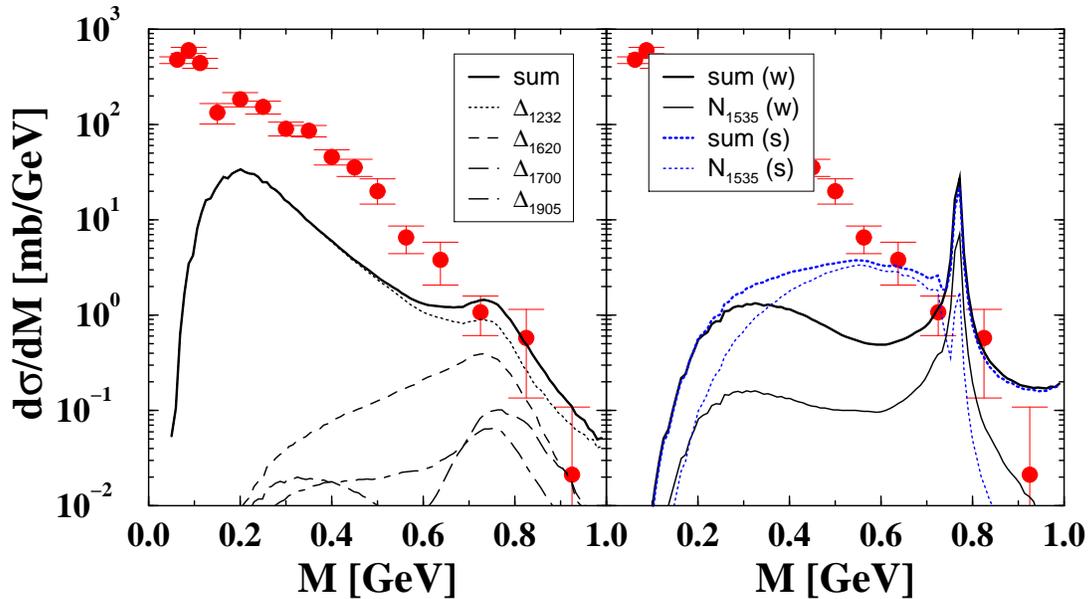}
\end{center}
\caption{Contributions of various nuclear resonances to the 
dilepton spectra in $Ca+Ca$ reactions at 1.04 AGeV. Left: 
contributions from $\Delta$ decays. Right: The total contribution 
from $N^*$ decays and that of the $N^*(1535)$ are shown for 
the two scenarios of a strong/weak (s/w) $N^*(1535) N\omega $ 
coupling. The DLS data \protect\cite{DLS} are shown in order to 
guide the eye.
}
\label{Cachan_fig}
\end{figure}
The contributions of the various nuclear resonances are displayed 
in Fig. \ref{Cachan_fig} for the Ca+Ca reaction.  
Here the theoretical results are not averaged according to the 
experimental resolution, but the DLS filter is applied and 
the data are also shown in order to guide the eye. The contributions from 
the $\Delta$ resonances which run exclusively over $\rho$ decays are 
dominated by the $\Delta (1232)$. However, in the vicinity of the 
$\rho$ peak the $\Delta (1620)$ gives an almost comparable contribution. 
The $\Delta (1700)$ and $\Delta (1905)$ give only minor contributions. 
The $N^*$ resonances which contribute both, via $\rho$ and $\omega$ decays 
are in particular important at invariant masses 
around and slightly below the $\rho/\omega$ peak. Before smearing over 
the experimental resolution the $\omega$ peak is clearly visible. 
As discussed in Sec. II in connection with the elementary production 
cross sections the $N^*(1535)$ plays a crucial role in our treatment. 
Therefore we display the contribution from this resonance separately 
for the two scenarios of a strong and a weak $N^*(1535) N\omega $ 
coupling. The first case (strong coupling) results in a smaller on-shell 
$\omega$ cross section which is reflected in a lower $\omega$ peak 
in the dilepton spectrum. The reason for the smaller on-shell value is 
a suppression of the $\omega$ strength from this resonance just 
at the $\omega$ pole \cite{omega02}. However, this scenario leads 
to a strong background contribution which is experimentally not 
accessible in $\omega$ production measurements but is clearly reflected 
in the enhanced dilepton spectrum below the $\omega$ pole. Compared 
to the weak coupling scenario the dilepton yield from $N^*(1535)$ is 
enhanced by almost one order of magnitude in this mass region. In 
the weak coupling scenario, on the other hand, the $N^*(1535)$ plays only 
a minor role in this kinematical region. 
\begin{figure}[h]
\begin{center}
\leavevmode
\epsfxsize = 9cm
\epsffile[60 40 430 380]{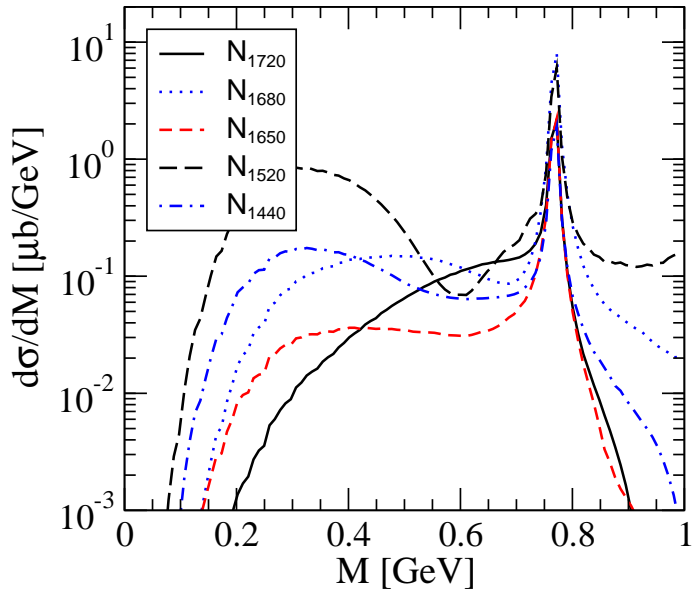}
\end{center}
\caption{Contributions of various $N^*$ resonances to the 
dilepton spectra in $Ca+Ca$ reactions at 1.04 AGeV. 
}
\label{CaCa_nres_fig}
\end{figure}
The contributions of the other $N^*$ resonances are shown in 
Fig. \ref{CaCa_nres_fig}. In the low mass region the most important 
one is the $N^*(1520)$ which has a strong $\rho$ decay mode \cite{krivo02}.
At the $\omega$ peak the $N^*(1520)$ and the $N^*(1680)$ dominate. 
Similar relative yields are obtained in $C+C$ reactions.

In summary one can conclude that the theoretical calculations without 
medium effects show in two distinct kinematical areas clear 
deviations from experiment: the low mass region between 
$M=0.1\div 0.5$ GeV is underestimated while the contribution at 
the $\omega$ (and $\rho$) peak is strongly overestimated. We investigated 
also the contributions from $\pi^+\pi^-$ annihilation. In our 
calculations the influence of this channel is significantly 
smaller than in \cite{BCRW98} and does not play an important role. 
\subsection{$\rho$- and $\omega$-meson in-medium widths}
In previous studies in-medium spectral functions of the $\rho$- and 
$\omega$-mesons were implemented into heavy-ion codes 
{\it ab initio} \cite{BCRW98}.  
At intermediate energies, the sensitivity of the dilepton spectra on 
the in-medium $\rho$-meson broadening is less pronounced as compared to
the $\omega$-meson. Estimates for the collision broadening 
of the $\rho$ in hadronic matter, i.e. dense nuclear matter or 
a hot pion gas, predict a collision width which is of the 
magnitude of the vacuum $\rho$ width. For the $\omega$, on the other hand, 
the vacuum width is only 8.4 MeV whereas in the medium it is 
expected to be more than one order of magnitude larger. However, 
the possibility of a strong in-medium modification of the $\omega$-meson 
has not attracted much attention in previous studies. The reason is 
probably due to the fact that the direct information on 
the $\omega$-meson channels from resonance decays, 
available through the multichannel $\pi N$ scattering analysis, 
is quite restricted. The present model provides an unified
description of the photo- and electro-production data and of the vector meson 
and dilepton decays of the nucleon resonances. 
It provides also a reasonable description of the vector meson 
and the dilepton production in elementary reactions ($p+p,p+d$) in 
the BEVALAC energy range. However, when applied to $A+A$ reactions 
the model leads to a very strong overestimation of the dilepton 
yield around the $\omega$-peak which suggests significant medium modifications 
of the $\omega$ contribution. At low energies, the vector 
meson production occurs due to decays of nucleon resonances.
The in-medium broadening of vector mesons can be understood within the 
framework of the resonance model. It has qualitatively two major 
consequences:
\begin{enumerate}
\item an increase of the nucleon resonance decay widths $R \rightarrow NV$ 

\item a decrease of the dilepton branchings $V \rightarrow e^+ e^-$  
due to the enhanced total vector meson widths. 
\end{enumerate}

These two effects are of opposite signs and can be 
completely described in terms of Eqs.(3)-(6) through 
appropriate modifications of the vector 
meson propagators entering into the $RN\gamma$ transition form 
factors $G_T(M^2)$. Within the eVMD framework it is sufficient to 
increase the total widths of the vector mesons. 
In a less formal way, the effect can be explained as follows: 
The differential branching 
\begin{equation}
dB(\mu,M)^{R\rightarrow NV} = 
\frac{d\Gamma_{\rm NV}^R (\mu,M)}{\Gamma_R (\mu)}
\label{bra}
\end{equation}
becomes usually larger with an increasing $V$ meson width which is 
due to the subthreshold character of the vector meson production through the 
light nucleon resonances. The dilepton branching of the nucleon resonances 
\begin{equation}
B(\mu)^{R\rightarrow Ne^+e^-} \sim
B(\mu)^{R\rightarrow NV} \frac{\Gamma_{V \rightarrow e^+e^-}}{\Gamma_V^{\rm tot}}
\label{bra1}
\end{equation}
is, on the other hand, inverse proportional to the total vector 
meson width $\Gamma_V^{\rm tot}$. Hence, 
an increase of the total width results in a decrease of the dilepton 
production rate. This effect is particularly strong for the 
$\omega$ since the in-medium $\omega$ width is expected to be
more than one order of magnitude greater than in the vacuum \cite{EBEK}. 
Although the estimates of ref. \cite{EBEK} were based on 
the standard VMD model which is contradictive with respect to the 
description of both, the $RNV$ and $RN\gamma$ branchings \cite{resdec,pirner,post01}, 
the qualitative conclusions concerning the magnitude of 
the in-medium $\omega$ broadening should be valid. 
A relatively large $\omega$ collision width is not too surprising. According to the 
$SU(3)$ symmetry the $\omega$ coupling to nucleons is 3 times greater than the $\rho$ 
coupling. One can therefore expect that 
at identical kinematical conditions the $N\omega$ cross section 
will be greater than the $N\rho$ cross section. Since the collision 
widths are proportional to the cross sections, the same 
conclusion holds for the collision widths.
The $\omega$ contribution is extremely sensitive to the reaction 
conditions in the course of the heavy ion collisions. While the 
increase of the total branching $B(\mu)^{R\rightarrow NV}$ depends 
on kinematical details one can expect that the suppression of the 
$\omega$ contribution due the enhanced total width $\Gamma_{\omega}^{\rm tot}$ 
is an one order of magnitude effect.
\begin{figure}
\begin{center}
\leavevmode
\epsfxsize = 10cm
\epsffile[20 40 700 580]{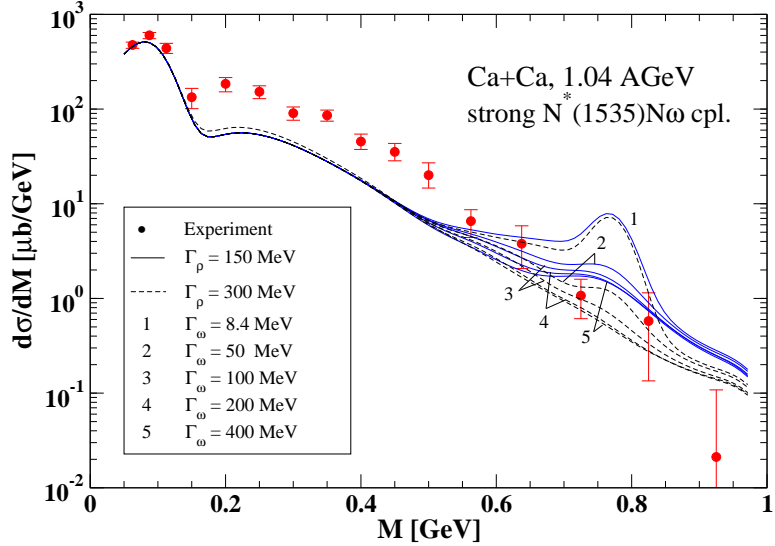}
\end{center}
\caption{Dilepton spectra in $Ca + Ca$ collisions at 1.04 AGeV for different 
values of the in-medium $\rho$ and $\omega$ widths.
The solid curves correspond calculations where the $\rho$ width is 
kept at its vacuum value of $150$ MeV (no collision broadening). 
The dashed curves correspond to a total $\rho$ width of 300 MeV. 
In both cases the $\omega$ width is varied between  
$\Gamma_{\omega}^{tot} = 8.4\div 400$ MeV. The results are obtained 
with the strong $N^*(1535)N\omega$ coupling. 
}
\label{DLS_fig3}
\end{figure}
In the standard approach without additional medium effects, 
Fig. \ref{DLS_AA_fig1}, 
both possibilities, i.e. the strong and the weak the $N^*(1535)N\omega$ 
decay mode, lead to a significant overestimation of the DLS data in the 
vicinity of the $\omega$ peak. An empirical way to investigate the influence 
of the collisional broadening is to assume in a first step 
average in-medium values for $\Gamma_{\rho/\omega}^{\rm tot}$ and 
to compare the corresponding results 
to the experiment. In Figs. \ref{DLS_fig3} and \ref{DLS_fig4} 
this is done for the $Ca+Ca$ reaction. The QMD results are shown 
for two values of the in-medium $\rho$ width, i.e. the vacuum 
value of 150 MeV and $\Gamma_{\rho}^{\rm tot}=300$ MeV. 
\begin{figure}
\begin{center}
\leavevmode
\epsfxsize = 10cm
\epsffile[20 40 700 580]{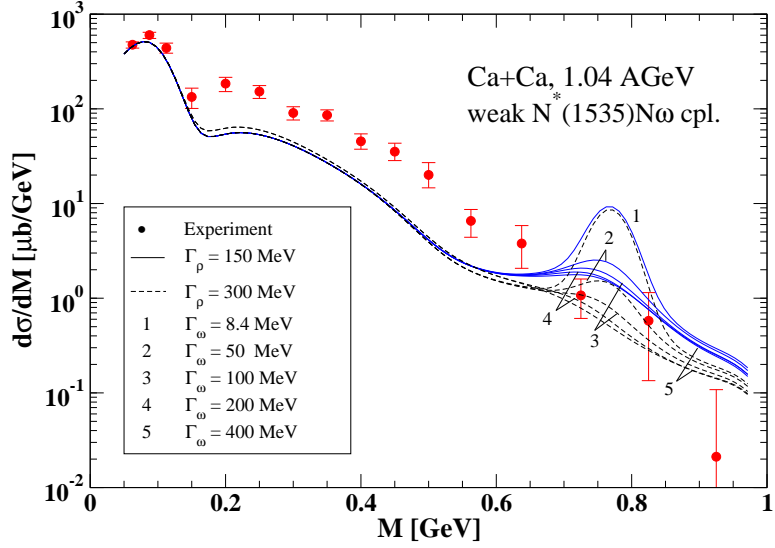}
\end{center}
\caption{Same as Figure \protect\ref{DLS_fig3}, but with 
weak $N^*(1535)N\omega$ coupling.
} 
\label{DLS_fig4}
\end{figure}
\begin{figure}
\begin{center}
\leavevmode
\epsfxsize = 10cm
\epsffile[20 40 700 580]{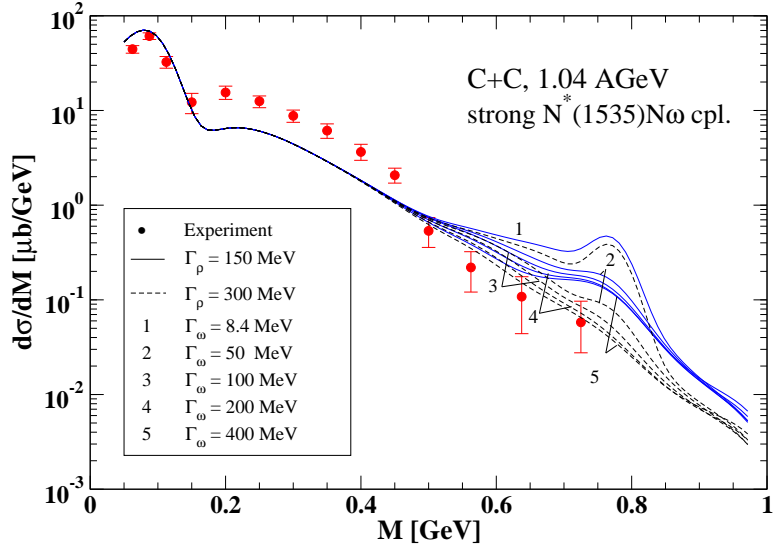}
\end{center}
\caption{Dilepton spectra in $C + C$ collisions at 1.04 AGeV for different 
values of the in-medium $\rho$ and $\omega$ widths.
The solid curves correspond calculations where the $\rho$ width is 
kept at its vacuum value of $150$ MeV (no collision broadening). 
The dashed curves correspond to a total $\rho$ width of $300$ MeV. 
In both cases the $\omega$ width is varied between  
$\Gamma_{\omega}^{tot} = 8.4\div 400$ MeV. The results are obtained 
with the strong $N^*(1535)N\omega$ coupling.
} 
\label{DLS_fig5}
\end{figure}
The latter assumes 
an additional collision width of $\Gamma_{\rho}^{\rm coll} =150$ MeV 
which agrees with the estimates of refs. \cite{KKW,EIoffe,EBEK,KS}. 
In both cases the $\omega$ 
width is varied between $\Gamma_{\omega}^{\rm tot}$ = 8.4, 50, 100, 200, and 400 MeV. 
As already mentioned, the in-medium $\omega$ broadening is 
less studied. Thus we cover the possible range of in-medium 
values by the above parameter set.

First of all, it is important to realize that the region which is 
sensitive to in-medium modifications of the meson widths is distinct 
from the mass interval between $0.2\div 0.6$ GeV where the DLS puzzle is observed. 
This means that the problem to extract in-medium vector meson widths is 
isolated from the difficulties concerning the theoretical interpretation of the 
dilepton spectra below the $\rho/\omega$ peak. As expected, the dilepton spectra 
in the vicinity of the $\rho/\omega$ peak react very sensitive on modifications 
of the in-medium width. The reproduction of the DLS data requires an 
in-medium $\omega$ width which lies above 50 MeV for both, strong and weak 
couplings. The best fits are obtained with $\Gamma_{\rho}^{\rm tot}$ = 300 MeV
and $\Gamma_{\omega}^{\rm tot}= 100\div 300$ MeV. With these values we 
reproduce in the strong $N^*(1535)N\omega$ 
coupling scenario the DLS data points around 
and 100 MeV below the $\rho/\omega$ peak within error bars. 
In the weak coupling scenario the DLS data are still slightly 
underestimated below the peak. However, the situation is not completely 
conclusive if one considers also the $C+C$ system, Fig. \ref{DLS_fig5}, 
where the strong coupling lies slightly above error bars even with in-medium 
meson widths. Definite conclusions on the $N^*(1535)N\omega$ mode 
from dilepton yields in heavy ion reactions require more precise data 
which will be provided by HADES \cite{friese}. The present estimates 
can be interpreted as empirical values which 
are directly extracted from the experiment. The strength of the $\omega$ broadening
and the theoretical motivation through Eq. (\ref{bra1}) provide 
confidence for these estimates. 

If the average widths are fixed one can, on the other hand, extract 
an average cross section from the collision broadening condition 
$\Gamma^{\rm coll}_{VN} = \langle \rho_B\rangle  v\gamma \sigma_{VN}$. 
The average nuclear density at the vector meson production, respectively at the 
decay of the corresponding nuclear resonances $R$, is in minimal 
bias 1 AGeV $Ca+Ca$ reactions about 1.5 times the saturation density, i.e. 
$\langle \rho_B\rangle_{Ca+Ca} = 0.24~{\rm fm}^{-3}$ and slightly less for 
$C+C$ ($\langle \rho_B\rangle_{C+C} = 0.20~{\rm fm}^{-3}$). If one assumes 
now that the vector mesons are produced in an isotropic fireball with 
a temperature of $T\simeq 80$ MeV the extracted collisional width corresponds 
to an average  $\rho N$ cross section of about $\sigma_{\rho N}\simeq 30$ mb 
and $\sigma_{\omega N}\simeq 50$ mb for the $\omega$ 
($\Gamma^{\rm tot}_{\omega}=200$ MeV).
\subsection{Decoherence}

The collision broadening of the vector mesons discussed above 
is most pronounced at invariant masses
close to $\rho$ and $\omega$ pole mass. A possible decoherence 
between the intermediate mesonic states in the resonance decays,  
in contrast, affects the dilepton spectrum preferentially 
below the $\rho/\omega$ peak 
( see Sec. III). The values which have already been extracted for 
the collision broadening of the vector mesons will therefore not 
significantly be changed when decoherence effects are additionally 
taken into account. Hence, we consider the values 
$\Gamma_{\rho}^{\rm coll} = 150$ MeV and 
$\Gamma_{\omega}^{\rm coll} = 100 - 300$ MeV already as final estimates which 
must not be iterated. 

The decoherence effect is treated as described in Sect. III. 
The collision broadening and the collision 
length are related through equations
\begin{equation}
e^{-l_C/L_C} = e^{- v t/L_C} = e^{-\Gamma^{\rm coll}_{V}t/\gamma }~~.
\label{reminder}
\end{equation}
Expression (\ref{reminder}) provides the the probability that 
a meson $V$ travels after its creation 
the length  $l_C$ through the medium without being scattered by 
the surrounding hadrons. In Eq.(\ref{reminder}), $v$ is velocity and $\gamma$
is the Lorentz factor. The collision length and width are thus related by
\begin{equation}
v /L_C = \Gamma^{coll}_{V}/\gamma ~~.
\label{reminder2}
\end{equation}
The collision length for the mesons is given by Eq.(\ref{collH}). 
An effective cross sections $\sigma_{VN}$ which is related to the 
collision width corresponds to Eq.(\ref{collH}), i.e. the factors 
$(1 + \alpha)e^{-\alpha}$ in Eq. (\ref{collH}) are then effectively included. 
Since the collision widths are directly extracted from data, 
the $\rho$ and $\omega$ collision lengths which are necessary in order 
to determine the probabilities for a coherent dilepton emission 
can be obtained from (\ref{reminder2}). The estimates of the collision lengths 
for radially excited vector mesons are thereby assumed to be the same as 
for the ground-state vector mesons. The vacuum widths
of the radially excited mesons are larger than those of the ground 
state $\rho$ and $\omega$. As a consequence, the radially excited mesons 
show a tendency to decay coherently. The 
decoherence effect is most pronounced for the ground-state $\omega$-meson, 
since its vacuum width is particularly small. The $\omega$-meson decays 
in the medium almost fully decoherent, i.e. after its first collision 
with another hadron. This results in 
a modification of the $N^* \rightarrow Ne^+e^-$ decay rates of the 
$I=1/2$ resonances due to the destruction of the interference 
between the $I=0$ and $I=1$ transition form factors. Since for the 
considered reactions the matter is isospin symmetric, 
the break up of the $\rho-\omega$ coherence does not result 
in a significant change of the dilepton spectra. In this case 
the isoscalar-isovector interference terms cancel on average. 
The major effect arises from the break up of the interference 
between the $\omega$ and its radial excitations.

\begin{figure}[h]
\begin{center}
\leavevmode
\epsfxsize = 14cm
\epsffile[20 50 550 390]{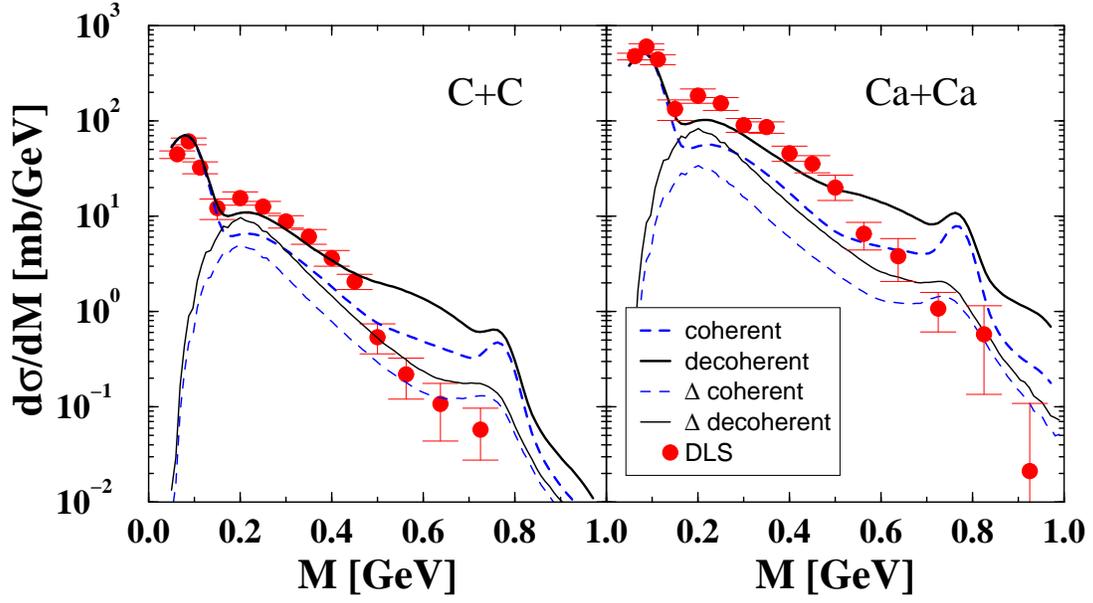}
\end{center}
\caption{Influence of a totally decoherent dilepton emission 
in $C+C$ and $Ca+Ca$ reactions. The contributions from the 
$\Delta$ resonances are in both cases shown separately.  
}
\label{DLS_AA_dec1}
\end{figure}

In Fig. \ref{DLS_AA_dec1} the influence of the decoherent 
summation of the intermediate 
mesonic states in the transition form factors is shown for both, 
the $Ca+Ca$ and $C+C$ reactions. 
To demonstrate the maximal possible effect we assume first total 
decoherence of all intermediate mesons. In this calculation no further 
medium effects are considered, i.e. the $\rho/\omega$ vacuum widths are 
used and the strong $N^*(1535)N\omega$ coupling 
is applied (the corresponding coherent calculations are the same as in 
Fig. \ref{DLS_AA_fig1}). A totally decoherent summation of the 
mesonic amplitudes in the 
resonance decays enhances the dilepton yield generally by about a factor 
of two. In the low mass region this enhancement is able to describe the 
DLS data. As can be seen from Fig. \ref{DLS_AA_dec1} this fact is 
due the enhancement of the $\Delta$ contributions by a factor of 2-3. 
However, also at larger invariant masses above 0.4 GeV the 
yield is enhanced and the spectrum is now stronger overestimated than 
in the coherent case. In the mass region between $0.4\div 0.8$ GeV 
the $N^*$ resonances give the major contribution to the yield. 
One has to keep in mind that the enhancement arises 
from the sum over the various $\Delta$ and $N^*$ resonances and the 
interplay between the corresponding electric, magnetic and Coulomb form 
factors. The enhancement is thus a complex function of the dilepton mass 
$M$. However, the scenario of a completely decoherent dilepton emission 
is rather unrealistic.
\begin{figure}[h]
\begin{center}
\leavevmode
\epsfxsize = 14cm
\epsffile[20 50 550 390]{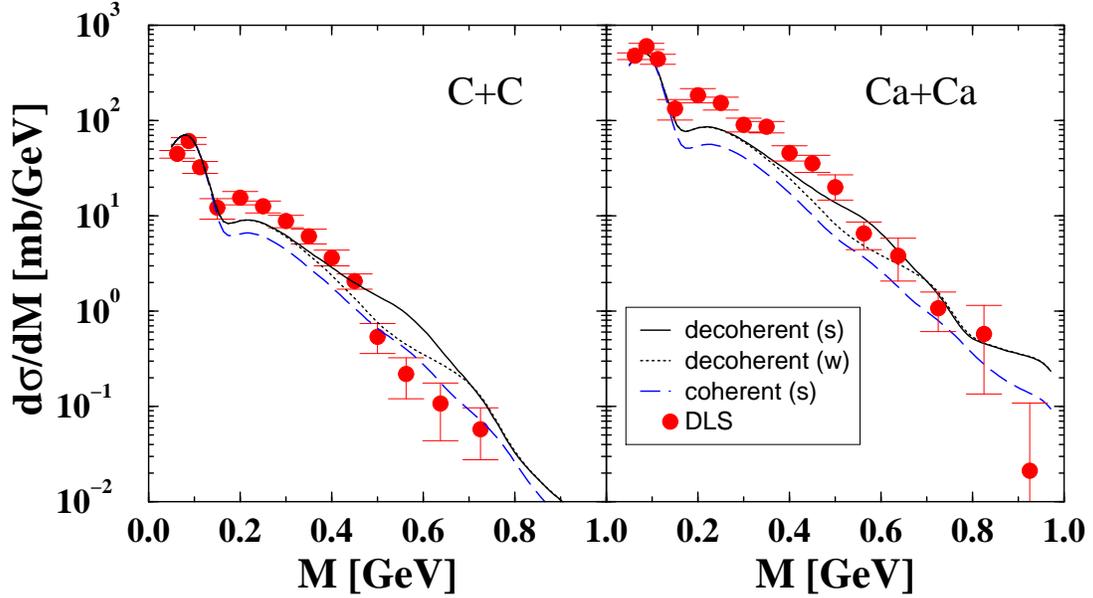}
\end{center}
\caption{Influence of the microscopically determined decoherent dilepton emission 
in $C+C$ and $Ca+Ca$ reactions. The calculations are 
performed with in-medium $\rho$ and $\omega$ widths of 300 and 200 MeV, 
respectively. The strong (s), respectively, weak (w) 
$N^*(1535)-N\omega $ coupling is used. For comparison also the 
coherent case (s) is shown. 
}
\label{DLS_AA_dec2}
\end{figure}
In a realistic calculation shown in Fig. \ref{DLS_AA_dec2} the 
probabilities for coherent/decoherent dilepton 
emission are determined microscopically as outlined above, i.e. by 
the use of Eqs. (\ref{decprob1}-\ref{coh3},\ref{reminder}). These 
realistic calculation are performed using the 'optimal' values for the 
in-medium widths of $\Gamma_{\rho}^{\rm coll}=150, 
\Gamma_{\omega}^{\rm coll}=200$ MeV. The low mass dilepton yield 
is now enhanced by about 50\% by the decoherence effect which is, 
however, still too less to describe the DLS data. The interplay between 
the two in-medium effects, i.e. the collisional broadening and the 
decoherent dilepton emission is more complex. Decoherence leads also 
to an enhancement of the dilepton yield in the mass region between 
$0.4\div0.7$ GeV. Since the main decoherence effect occurs through 
the broken interference of the $\omega$ with its excited states, 
it is most pronounced in the dilepton contribution which stems from the 
$N^*$ resonance decays. This explains the difference between the 
two calculations assuming a strong/weak $N^*(1535)N\omega$ coupling 
in the mass range where possible off-shell $\omega$ contributions are 
now enhanced (strong coupling). However, definite conclusions on the 
strength of the $N^*(1535)N\omega$ coupling are still difficult to 
make at the present data situation. For the strong coupling the 
$Ca+Ca$ system is in agreement 
within error bars with the DLS data whereas in the lighter $C+C$ 
system the data are now overestimated and would favor the weak coupling. 
In both cases the agreement with the data is significantly improved in 
the low mass region. However, the considered decoherence 
effects are not completely sufficient in order to solve the DLS puzzle.  
The reason is that the microscopic determination of the decoherence 
probability favors the break up of the coherence between the $\omega$ 
and its excited states in the $N^*$ decays rather than the 
break up between the $\rho$ and its excited states in the $\Delta$ decays. 
The latter resonances are, however, those which contribute to most extent 
at low invariant masses. 
\section{Conclusion}
In the present work we provided a systematic description of vector 
meson and dilepton production in elementary $NN$ and $\pi N$ as well 
as in $A+A$ reactions. The reactions dynamics of the heavy ion collisions 
is described by the QMD transport model which was extended for the inclusion of 
nucleon resonances with masses up to 2 GeV. The vector meson production in 
elementary reactions is described through excitations of nuclear resonances 
within the framework of an extended VMD model. The model parameters were 
fixed utilizing electro- and photo-production data as well 
as $\pi N$ scattering analysis. Available data on the $\rho$ and $\omega$ 
production in $p+p$ and $\pi+N$ reactions are well reproduced. The same holds for 
the dilepton production in elementary $p+p$ and $p+d$ reactions. 

The situation becomes different turning to heavy ion collisions: 
In $C+C$ and $Ca+Ca$ reactions we observe in two distinct 
kinematical regions significant deviations from 
the dilepton yields measured by the DLS Collaboration. At small 
invariant masses the experimental data 
are strongly underestimated which confirms the observations made 
by other groups. Although accounting for the experimental resolution 
we observe further a clear structure of the $\rho/\omega$ peak which is  
not present in the data. Both features imply the investigation 
of further medium effects. 

The collisional broadening of the vector mesons suppresses the  
$\rho/\omega$ peak in the dilepton spectra. This allows to extract  
empirical values for the in-medium widths of the vector mesons. From  
the reproduction of the DLS data the following estimates for the 
collision widths $\Gamma_{\rho}^{\rm coll} = 150$ MeV and 
$\Gamma_{\omega}^{\rm coll} = 100 - 300$ MeV can be made. The in-medium values 
correspond to an average nuclear density of about 1.5 $\rho_0$. HADES will 
certainly help to constrain these values with higher precision. 

The second medium effect discussed here concerns the problem of 
quantum interference. Semi-classical transport models like QMD do generally 
not account for interference effects, i.e. they propagate probabilities 
rather than amplitudes and assume that relative phases cancel the 
interference on average. 
However, interference effects can play an important role for the dilepton 
production. In the present model the decay of nuclear resonances which is 
the dominant source for the dilepton yield, requires the destructive interference 
of intermediate $\rho$ and $\omega$ mesons with their excited states. 
The interference can at least partially be destroyed by the presence of 
the medium which leads to an enhancement of the corresponding dilepton 
yield. We proposed a scheme to treat the decoherence in the medium on 
a microscopic level. The account for decoherence improves the agreement with 
the DLS data in the low mass region. However, the magnitude of this effect is 
not sufficient to resolve the DLS puzzle completely. \\
\noindent
{\bf Acknowledgements}\\
This work was supported by the BMBF under contract 
06T\"U986, the DFG under grant 436RUS 113/721/0 and RFBR under 
grant 03-02-04004.



\newpage
\begin{table}
\begin{center}
\caption{Coefficients for the isotopic decomposition of 
the $ NN\rightarrow NN\rho$ cross section into contributions 
from $\Delta$ and $N^*$ resonances.}
\begin{tabular}{ccc}
                              & $\alpha$  & $\beta$ \\
\hline
$pp \rightarrow pp\rho^0$     & 1/6 & 1/3 \\
$pp \rightarrow pn\rho^+$     & 5/6 & 2/3 \\
$nn \rightarrow nn\rho^0$     & 1/6 & 1/3 \\
$nn \rightarrow np\rho^-$     & 5/6 & 2/3 \\
$np \rightarrow np\rho^0$     & 1/3 & 1/3 \\
$np \rightarrow pp\rho^-$     & 1/12 & 1/3 \\
$np \rightarrow nn\rho^+$     & 1/12 & 1/3 \\
\end{tabular}
\label{iso_tab1}
\end{center}
\end{table}

\begin{table}
\begin{center}
\caption{Coefficients for the isotopic decomposition of
the $ \pi N\rightarrow \rho N$ cross section into contributions
from $\Delta$ and $N^*$ resonances.}
\begin{tabular}{ccc}
                              & $\alpha$  & $\beta$ \\
\hline
$\pi^+ p \rightarrow \rho^+ p$     &  1  &  0  \\
$\pi^+ n \rightarrow \rho^+ n$     & 1/9 & 4/9 \\
$\pi^+ n \rightarrow \rho^0 p$     & 2/9 & 2/9 \\
$\pi^0 p \rightarrow \rho^+ n$     & 2/9 & 2/9 \\
$\pi^0 p \rightarrow \rho^0 p$     & 4/9 & 1/9 \\
$\pi^0 n \rightarrow \rho^0 n$     & 4/9 & 1/9 \\
$\pi^0 n \rightarrow \rho^- p$     & 2/9 & 2/9 \\
$\pi^- p \rightarrow \rho^0 n$     & 2/9 & 2/9 \\
$\pi^- p \rightarrow \rho^- p$     & 1/9 & 4/9 \\
$\pi^- n \rightarrow \rho^- n$     &  1  &  0  \\
\end{tabular}
\label{iso_tab2}
\end{center}
\end{table}

\begin{table}
\begin{center}
\caption{List of $N^*$ resonances which are included in 
the QMD transport model. The table shows the resonances masses and the 
total and partial widths of the included decay channels in MeV. 
The values of $\Gamma_{N\omega}$ and $\Gamma_{N\rho}$ are given 
at the resonance pole masses. The values in brackets as well as the 
other decay channels are taken from 
\protect\cite{Bass} and used for the reaction dynamics. }
\begin{tabular}{cccccccccc}
Res. & Mass [MeV] & $\Gamma_{\rm tot}$ [MeV] & $N\omega$  & $N\rho$  & $N\pi$ & $N\pi\pi$
& $\D_{1232}\pi$ & $N_{1440}\pi$ & $N\eta$ \\
\hline
$N_{1440}$ & 1440 & 200 & \makebox[1cm][l]{$< 10^{-4}$} \makebox[1cm][r]{(--)}
& \makebox[1cm][l]{0.45} \makebox[1cm][r]{(--)}  &  140 &    10 &    50 &    -- &    --\\
$N_{1520}$ & 1520 & 125 & \makebox[1cm][l]{0.08} \makebox[1cm][r]{(--)}
& \makebox[1cm][l]{26.63} \makebox[1cm][r]{(--)} &   75 & 18.75 & 31.25 &    -- &    --\\
$N_{1535}$ & 1535 & 150 & \makebox[1cm][l]{2.05} \makebox[1cm][r]{(--)}
& \makebox[1cm][l]{4.62} \makebox[1cm][r]{(--)}  & 82.5 &   7.5 &     - &  7.5 & 52.5\\
$N_{1650}$ & 1650 & 150 & \makebox[1cm][l]{0.94} \makebox[1cm][r]{(--)}
& \makebox[1cm][l]{3.17} \makebox[1cm][r]{(--)}  & 97.5 &   7.5 &    15 &  7.5 &  7.5\\
$N_{1675}$ & 1675 & 140 & \makebox[1cm][l]{0.003} \makebox[1cm][r]{(--)}
& \makebox[1cm][l]{3.50} \makebox[1cm][r]{(--)} &   63 &    77 &     -- &    -- &    --\\
$N_{1680}$ & 1680 & 120 & \makebox[1cm][l]{0.50} \makebox[1cm][r]{(--)}
& \makebox[1cm][l]{10.24} \makebox[1cm][r]{(24)} &   78 &    18 &     -- &    -- &    --\\
$N_{1700}$ & 1700 & 100 & \makebox[1cm][l]{--} \makebox[1cm][r]{(--)}
& \makebox[1cm][l]{--} \makebox[1cm][r]{(5)} &   10 &    45 &    35 &    -- &    5\\
$N_{1710}$ & 1710 & 110 & \makebox[1cm][l]{--} \makebox[1cm][r]{(--)}
& \makebox[1cm][l]{--} \makebox[1cm][r]{(5.5)} & 16.5 &    22 &    22 &   11 &   22\\
$N_{1720}$ & 1720 & \makebox[0.8cm][l]{184} \makebox[0.8cm][r]{(150)} 
& \makebox[1cm][l]{32.4} \makebox[1cm][r]{(--)} &
\makebox[1cm][l]{129.3} \makebox[1cm][r]{(37.5)} & 22.5 &  67.5 &    15 &    -- &    --\\
$N_{1900}$ & 1870 & 500 & \makebox[1cm][l]{--} \makebox[1cm][r]{(275)}
& \makebox[1cm][l]{--} \makebox[1cm][r]{(25)} &  175 &     -- &    25 &    -- &    --\\
$N_{1990}$ & 1990 & 550 & \makebox[1cm][l]{--} \makebox[1cm][r]{(--)}
& \makebox[1cm][l]{--} \makebox[1cm][r]{(82.5)} & 27.5 & 137.5 &   165 & 82.5 &    --\\
\end{tabular}
\label{nstar_tab}
\end{center}
\end{table}

\begin{table}
\begin{center}
\caption{List of $\Delta$ resonances which are included in 
the QMD transport model. The table shows the resonances masses and the 
total and partial widths of the included decay channels in MeV. 
The values of $\Gamma_{N\rho}$ are given 
at the resonance pole masses. The values in brackets as well as the 
other decay channels are taken from 
\protect\cite{Bass} and used for the reaction dynamics. }
\begin{tabular}{ccccccc}
Res. & Mass [MeV] & $\Gamma_{\rm tot}$ [MeV] & $N\rho$  & $N\pi$ & $\D_{1232}\pi$ & $N_{1440}\pi$\\
\hline
$\D_{1232}$ & 1232 & 115 & \makebox[1cm][l]{$\sim 0$}
\footnote{At the resonance pole $\Gamma_{N\rho}$ is practically zero for 
the $\D_{1232}$ due to vanishing phase space. However, the $\rho$-meson 
coupling constants of this resonance, in particular 
the magnetic one, are large \protect\cite{krivo02} and thus the 
$\D_{1232}$ has non-vanishing off-shell contributions.}       

\makebox[1cm][r]{(--)}    &   115 &    -- &    --\\
$\D_{1600}$ & 1700 & 200 & \makebox[1cm][l]{--}   \makebox[1cm][r]{(--)}    &    30 &  110 &   60\\
$\D_{1620}$ & 1675 & 180 & \makebox[1cm][l]{16.4}  \makebox[1cm][r]{(--)}    &    45 &  108 &   27\\
$\D_{1700}$ & 1750 & 300 & \makebox[1cm][l]{47.7}  \makebox[1cm][r]{(30)}   &    60 &  165 &   45\\
$\D_{1900}$ & 1850 & 240 & \makebox[1cm][l]{--}  \makebox[1cm][r]{(36)}   &    72 &   72 &   60\\
$\D_{1905}$ & 1880 & \makebox[0.8cm][l]{363} \makebox[0.8cm][r]{(280)} & 
\makebox[1cm][l]{307.3} \makebox[1cm][r]{(168)}  &    56 &   28 &   28\\
$\D_{1910}$ & 1900 & 250 & \makebox[1cm][l]{--}  \makebox[1cm][r]{(100)}  &  87.5 & 37.5 &   25\\
$\D_{1920}$ & 1920 & 150 & \makebox[1cm][l]{--}  \makebox[1cm][r]{(45)}   &  22.5 &   45 & 37.5\\
$\D_{1930}$ & 1930 & 250 & \makebox[1cm][l]{--} \makebox[1cm][r]{(62.5)} &    50 & 62.5 &   75\\
$\D_{1950}$ & 1950 & 250 & \makebox[1cm][l]{--}   \makebox[1cm][r]{(37.5)} & 112.5 &   50 &   50\\
\end{tabular}
\label{delta_tab}
\end{center}
\end{table}

\end{document}